\documentclass{aa}
\usepackage{amssymb,amsmath}
\usepackage{natbib}
\usepackage{txfonts}
\bibpunct{(}{)}{;}{a}{}{,}%
\usepackage[dvips]{graphicx}
\usepackage{color}

\makeatletter
  \def\mathch{\protect\p@mathch}
  \def\p@mathch#1#2{%
    \begingroup
    \let\@nomath\@gobble \mathversion{#1}%
    \math@atom{#2}{%
    \mathchoice%
      {\hbox{$\m@th\displaystyle#2$}}%
      {\hbox{$\m@th\textstyle#2$}}%
      {\hbox{$\m@th\scriptstyle#2$}}%
      {\hbox{$\m@th\scriptscriptstyle#2$}}}%
  \endgroup}
  \def\bmath{\protect\p@boldm}
  \def\p@boldm#1{\mathch{bold}{#1}}
\makeatother
\newcommand\bF{\vec{F}}
\newcommand\bG{\vec{G}}

\newcommand\be{\vec{e}}

\newcommand\bu{\vec{u}}

\newcommand\bnabla{\vec{\nabla}}

\newcommand\p{\partial}

\newcommand\rmd{\mathrm{d}}

\newcommand\rmp{\mathrm{p}}
\newcommand\rmc{\mathrm{c}}

\newcommand\twod{{\mbox{\scriptsize 2D}}}
\newcommand\threed{{\mbox{\scriptsize 3D}}}

\newcommand{\grad}{\vec{\nabla}}
\newcommand{\curl}{\vec{\nabla}\times}
\renewcommand{\div}{\vec{\nabla}\cdot}

\newcommand{\f}[2]{\frac{#1}{#2}}

\newcommand{\tay}{\mathrm{Ta}}
\newcommand{\rey}{\mathrm{Re}}

\newcommand{\rom}{R_\Omega}

\newcommand{\difft}[1]{\frac{\partial #1}{\partial t}}

\newcommand{\diffy}[1]{\frac{\partial #1}{\partial y}}
\newcommand{\diffz}[1]{\frac{\partial #1}{\partial z}}

\newcommand{\ddiffyy}[1]{\frac{\partial^2 #1}{\partial y^2}}

\title{On self-sustaining processes in Rayleigh-stable rotating\\ plane Couette
  flows and subcritical transition to\\ turbulence in accretion disks}
\titlerunning{Self-sustaining processes in Rayleigh-stable rotating plane Couette flows}

\author{F. Rincon\inst{1} \and G. I. Ogilvie\inst{1} \and C. Cossu\inst{2}}
\authorrunning{F. Rincon, G. I. Ogilvie \and C. Cossu}

\institute{Department of Applied Mathematics and Theoretical Physics,
  University of Cambridge, Centre for Mathematical Sciences, Wilberforce
  Road, Cambridge CB3 0WA, United Kingdom\\
  \email{F.Rincon@damtp.cam.ac.uk} \and Laboratoire d'hydrodynamique (LadHyX), CNRS -- \'Ecole
   polytechnique, 91128 Palaiseau cedex, France}

\offprints{F.~Rincon}

\date{\today}

\begin{document}

\abstract
{Subcritical transition to turbulence in Keplerian accretion disks is
still a controversial issue and some theoretical progress is required 
in order to determine whether or not this scenario provides a plausible
explanation for the origin of  angular momentum transport in
non-magnetized accretion disks.}
{Motivated by the recent discoveries of
exact nonlinear steady self-sustaining solutions in linearly stable
non-rotating shear flows, we attempt to compute similar solutions in
Rayleigh-stable rotating plane Couette flows and to identify
transition mechanisms in such flows.}
{Numerical simulations (time-stepping approaches) have been
 employed in most recent studies of the problem. Here,
 we instead combine nonlinear continuation methods and asymptotic theory 
 to study the nonlinear dynamics of plane Couette flows in Rayleigh-stable regimes.}
{We obtain exact nonlinear solutions for Rayleigh-stable cyclonic
 regimes using continuation techniques proposed by Waleffe and Nagata 
 but show that it is not possible to compute solutions for Rayleigh-stable
 anticyclonic regimes, including Keplerian flow, using similar
 techniques. We also present asymptotic descriptions of these various
 problems at large Reynolds numbers that provide some insight into the
 differences between the non-rotating and Rayleigh-stable anticyclonic
 regimes and derive some necessary conditions for mechanisms analogous
 to the non-rotating self-sustaining process to be present in 
 (constant angular momentum) flows on the Rayleigh line.}
 {Our results demonstrate that subcritical transition mechanisms 
  cannot be identified in wall-bounded Rayleigh-stable anticyclonic
  shear flows by transposing directly the phenomenology of subcritical
  transition in cyclonic and non-rotating wall-bounded shear flows. 
  Asymptotic developments, however, leave open the possibility that
  nonlinear self-sustaining solutions may exist in unbounded or periodic
  flows on the Rayleigh line. These could serve as a starting point to
  discover solutions in Rayleigh-stable flows, but the nonlinear
  stability of Keplerian accretion disks remains to be determined.}

\maketitle

\section{Introduction}
Physical mechanisms responsible for turbulent angular momentum transport
in Keplerian accretion disks have long been a  matter of debate
among astrophysicists. While there is convincing evidence that
the presence of magnetic fields in disks \citep{donati05} triggers
turbulence thanks to a linear magnetorotational instability \cite[MRI,
see][]{velikhov59,chandra60,balbus91}, purely hydrodynamic
sources of turbulence still have to be discovered in some regions
of protoplanetary disks which are likely to be decoupled from magnetic fields
\citep{fromang02} and in disks forming in binary systems in quiescence, 
in which MHD turbulence and the associated turbulent transport
may die away as a consequence of the smallness of the magnetic Reynolds
number \citep{gammie98}.

Studying hydrodynamic transition to
turbulence in Keplerian flows is an extremely difficult task for
essentially two reasons. On the one hand, there is currently no
known local linear instability which could provide a route to turbulence
in the non-magnetized Keplerian flow regime. \cite{dubrulle05} have
pointed out the existence of an instability in stably stratified
accretion disks, but its local nature is very unclear
\citep{umurhan06,branden06}. On the other
hand, the linear operators associated with shear flows are
mathematically  non-normal, which physically means that
short-term dynamics (transient algebraic growth) rather than normal
modes may play an important role in the process of transition to turbulence
\citep{korycansky92,trefethen93,schmid00,chagelishvili03,tevzadze03,
yecko04,afshordi05,narayan05}.  Several authors \citep{richard99,
longaretti02} have suggested that subcritical transition  may be
possible in accretion
disks in spite of the stabilizing effect of the Coriolis force, 
but this point of view has been contested by \cite{balbus96}, \cite{hawley99} and
\cite{balbus03}. The relevance of such mechanisms to the problem of
turbulent transport in disks has also been questioned by \cite{lesur05}, who
extrapolated from numerical
simulation results that the critical Reynolds number for this kind of process
to occur in the Keplerian regime was far too large and the associated
angular momentum transport far too small to account for the accretion
process. Recently, \cite{balbus06} have further questioned this
transition scenario, basing their argument on the observation that
transiently amplified shearing waves are solutions for any amplitude
(because their self-interactions are vanishing) and cannot 
therefore directly trigger turbulence. \cite{balbus06} and \cite{shen06} have
also studied the nonlinear evolution of these non-axisymmetric shearing waves
using high-resolution simulations and have found  that  they become
unstable to Kelvin-Helmholtz instabilities if they reach a sufficient
amplitude. However, the turbulent motions generated this way ultimately
die away in the configuration of their simulation.


Most recent theoretical studies of hydrodynamic turbulence in accretion
disks have focused on linear transient growth, which is only one part of
a full  subcritical transition scenario, and is not a sufficient condition
for nonlinear instability.
Nonlinear dynamics has mostly been studied numerically within the shearing
box model \citep{rogallo81} using direct time-stepping approaches. The nonlinear
outcome of the transient growth process remains unclear and
theoretical progress is required in order to clarify the existence and nature
of subcritical transition in Rayleigh-stable (centrifugally stable)
rotating shear flows. The aim of this paper is to present some theoretical and numerical
arguments that may help to make progress in this direction, and notably
to investigate if some nonlinear mechanisms that are crucial for subcritical
transition to turbulence in non-rotating shear flows play a role in
linearly stable rotating shear flows too. We have been motivated by the
vast amount of experimental, numerical and theoretical work that has
become available in fluid dynamics in the last ten years and have 
oriented our study towards one of the most promising avenues of research in that
field, namely the self-sustaining process, first identified by
\cite{hamilton95} for non-rotating plane shear flows. Our hope has
been to be able to identify  a similar process for Rayleigh-stable
rotating shear flows, notably the anticyclonic plane Couette flow on the
Rayleigh line, which at first glance presented many similarities with its
non-rotating counterpart  \citep{balbus98}. 
Unlike most recent studies, we have tried to address the fully nonlinear
problem using continuation methods, in order to compute
accurate nonlinear solutions in different rotation regimes and to pinpoint the 
physical mechanisms that may give birth to turbulence in these regimes.
Our results shed some new light on the nonlinear outcome of linear transient
growth on the Rayleigh line and show that it is not possible to identify
nonlinear solutions in  wall-bounded Rayleigh-stable anticyclonic shear
flows using arguments similar to those pertaining to the self-sustaining
process in non-rotating shear flows. We also argue that these
findings do not rule out the possibility of a subcritical transition
scenario in accretion disks, but show that self-sustaining mechanisms in
rotating plane shear flows with linearly stable anticyclonic rotation,
if they exist, must be qualitatively different from their non-rotating equivalent.

The outline of the paper is as follows. In Sect.~\ref{review}, we  
summarize the current understanding of transition to turbulence in 
shear flows and give a detailed description of the phenomenology of the
self-sustaining process for non-rotating shear flows. Section~\ref{equations}
is devoted to the mathematical description of the nonlinear problem that
we have been attempting to solve. The numerical tools developed
for the purpose of our study are presented in Sect.~\ref{nummethods}.
We then report some  numerical results for subcritical transition in
cyclonic (Sect.~\ref{numerics_cycl}) and anticyclonic
(Sect.~\ref{numerics_anti})  rotating plane Couette flows. Section~\ref{theory}
provides some theoretical insight into
the asymptotics of self-sustaining processes at very large Reynolds
numbers in order to understand these numerical results and the difficulties
in identifying a self-sustaining process so far for linearly stable
anticyclonic regimes. A discussion of the results and guidelines for
future work are finally presented in Sect.~\ref{conclusions}.

\section{Transition to turbulence in shear flows and the self-sustaining
  process\label{review}}
To gain some theoretical understanding of the Keplerian problem, 
it is necessary to come back to the more general
problem of hydrodynamic transition to turbulence in shear
flows. For a long time, the only classical non-rotating shear flow for which an 
explanation for transition was available is plane Poiseuille flow with
rigid boundaries. The transition was attributed to the linear
instability of this flow at a critical Reynolds number $\rey=5572$
(see \citet{orszag71,zahn74,orszag80}), which takes the form of travelling
waves. The instability is subcritical and finite amplitude solutions
can be tracked down to $\rey=2900$ \citep{herbert76}. However, it was noticed that this
global critical Reynolds number was rather high compared to the
transition value observed experimentally. Studies of secondary instabilities of
these travelling waves did not seem to resolve this discrepancy
\citep[see \textit{e.~g.}][]{drissi99}. Another major
problem with this approach was  that it  applied only to plane
Poiseuille flow: plane Couette flow and pipe Poiseuille flow do
not have any linear instability and nevertheless become turbulent
at moderate values of $\rey$ (the linear stability for all
  Reynolds numbers of Plane Couette flow with rigid boundaries was
  demonstrated by \cite{romanov73}).

In the last decade,  a nonlinear self-sustaining process acting as an
organizing center of the dynamics in phase space has been discovered, leading to a
new paradigm in the understanding of transition in shear flows. Corresponding exact
nonlinear solutions to the Navier-Stokes equations have  been
identified numerically in plane Poiseuille flow and plane Couette flow
 with either no-slip or stress-free boundaries
\citep{hamilton95,waleffe95,waleffe97,schmiegel99,waleffe01,waleffe03},
 and in pipe Poiseuille flow \citep{faisst03,wedin04}, leading to an
 elegant solution to the historical problem of transition in pipes
 \citep{reynolds83}. These coherent
 structures take the form of nonlinear travelling waves or of a steady
 pattern, depending on the symmetries of the original problem. The
 critical global transition Reynolds numbers and
 the mean flow profiles for these solutions prove to be far
 closer to experimental measurements than predictions made by other
 theories, and an experimental detection of these structures in pipe
 Poiseuille flow has been reported recently \citep{Hof04}.
The self-sustaining process as it is understood now is a subtle balance
between linear and nonlinear mechanisms. The \textit{linear part} of the
process results physically from the transient growth of a
streamwise-independent (axisymmetric in the accretion disk terminology)
streamwise velocity field (referred to as the streaks field)
which originates from the redistribution of the mean shear
 by low-amplitude streamwise vortices (or equivalently streamwise
 rolls). Such a growth is limited to time scales $\mathcal{O}(\rey)$
 because of the ultimate diffusion of streamwise vortices. This
 mechanism is called the lift-up effect
 \citep{landahl80,butler92}.  A \textit{steady state} can only be
 achieved thanks to the development of a three-dimensional
 linear instability of the high-amplitude streaks, which produces a
 \textit{nonlinear feedback} that is sufficient to maintain the
 original rolls field. Note that the linear growth of
 streamwise-independent structures is a key requirement for the whole
 process to occur, and that this  physical mechanism is a direct
 consequence of the non-normality of the  linear operator. It is also
 different from and more powerful than the  ``swing amplification'' of
 streamwise-dependent (non-axisymmetric)  Orr-Kelvin shearing waves
 \citep{kelvin1887,orr1907,knobloch84,craik86,butler92}. It is worth
 emphasizing here that the upper and lower branches  of these three-dimensional
 solutions are both unstable to steady and oscillatory modes with the
same basic symmetries. As discussed by \cite{waleffe01}, this does not
mean that the original coherent structures are not relevant to
transition. On the contrary, they are expected to show up intermittently
in the flow, and the fact that they have instabilities ensures that
bifurcations to even more complex structures and to turbulence are
possible.

Far less is known analytically regarding the stability of spanwise 
rotating plane shear flows (the spanwise direction is the one
perpendicular to the disk plane in an accretion disk). 
The linear stability properties of such systems depend
on the rotation number $\rom$, which is proportional to the ratio
of the rotation rate $\Omega$ and the background shear rate $S$
(defined here as the negative of the background flow relative vorticity):
\begin{equation}
  \label{eq:rotation}
\rom=-\f{2\Omega}{S}.
\end{equation}
This $\rom$ is the same as the one defined by \cite{lesur05} ($S$ here is
$-S$ in their paper but they use $\rom=2\Omega/S$) and differs in sign
from the definition of \cite{bech97}.
Here we assume that $S$ is constant in the shearwise (radial) direction, 
which corresponds to a plane Couette flow. A necessary condition (known
as the Rayleigh criterion) for this flow to become linearly unstable is that
$-1<\rom<0$, which shows that the flow has to be anticyclonic to be unstable, 
\textit{i.~e.} that the vorticity of the background  flow and the
rotation rate have opposite sign. In this so-called Rayleigh-unstable
regime, the centrifugal instability is responsible for the transition to turbulence,
as  observed in both laboratory \citep{tillmark96,alfredsson05} and numerical
\citep{bech96,bech97} experiments. This instability is the same 
as the instability studied by \cite{taylor23} in the Taylor-Couette
experiment, which exhibits axisymmetric  structures called Taylor vortices
close to the bifurcation point. On the contrary, rotating plane Couette flow in the
cyclonic regime ($\rom> 0$) and anticyclonic flows with $\rom< -1$
are Rayleigh-stable for all $\rey$. Local linearizations of Keplerian
flows, which have $\rom=-4/3$, are included into this category.
From the experimental point of view, turbulence is of
course known to be present in plane Couette flow for $\rom=0$  (non-rotating plane
Couette flow), but it has also been observed for small positive rotation
numbers \citep{tillmark96}. Note finally that unlike for
  non-rotating plane Couette flow,  there is no complete proof that
  rotating plane Couette flow at  $\rom=-1$ is linearly stable for all
  Reynolds numbers.

At that point, it should be noted that a
strong theoretical connection exists between the nonlinear stability of
non-rotating plane Couette flow and the weakly nonlinear behaviour of 
Rayleigh-unstable Taylor-Couette flow in the narrow gap limit. This
link was discovered by \cite{nagata88,nagata90}. He found that the
three-dimensional nonlinear self-sustaining steady solutions for
non-rotating plane Couette flow (which were unknown at that time and had
therefore not been related to any kind of self-sustaining
process) could be obtained directly by nonlinear continuation (with
respect to the rotation rate) of wavy Taylor vortices, which emerge from
a three-dimensional bifurcation of the Taylor vortex flow. For a
given range of streamwise wavenumbers, \cite{nagata90} showed that these
solutions could  be continued slightly into the cyclonic regime. 
\cite{komminaho96} performed numerical simulations and discovered that
the large-scale structures of turbulence disappeared in the cyclonic
regime at  $\rom=0.06$ for $\rey=750$, in qualitative agreement
with the experimental results of \cite{tillmark96}.

Whether or not turbulence exists in the Rayleigh-stable,
anticyclonic regime (and notably in Keplerian accretion disks) is still
very unclear, as mentioned earlier.
\cite{richard99}, using data from experiments by \cite{wendt33} and \cite{taylor36},
suggested that transition to turbulence should be observed in a
Taylor-Couette flow in that regime at rather modest $\mathcal{O}(10^5)$
Reynolds numbers, and claimed  experimental evidence for this 
\citep{richard01}. \cite{bech97} found that turbulence disappeared 
at $\rom=-1$ for Reynolds numbers up to $\rey\sim 1300$. 
Nonlinear stability in the Taylor-Couette problem was also
investigated by \cite{garaud05} using a closure model for the Reynolds
stresses derived by \cite{ogilvie03}. They showed that within the
framework of this model, nonlinear solutions in the anticyclonic
Rayleigh-stable regime should exist if solutions in
the cyclonic regime exist, thanks to a symmetry with respect to
$\rom=-0.5$. Streamwise-independent solutions of the Navier-Stokes
equations do have such a symmetry \citep[see also][]{balbus96}, but \cite{lesur05}
pointed out that it is  broken when fully three-dimensional solutions of
the Taylor-Couette problem are considered, leaving the issue of the
existence of nonlinear solutions in the stable anticyclonic regime
unsettled.  \cite{hawley99} and \cite{lesur05}, using
shearing box simulations \citep{rogallo81}, observed turbulence 
for anticyclonic rotation down to $\rom=-1.03$ and \cite{lesur05},
using either shearing box or rigid-boundary simulations, found
turbulence at small positive rotation numbers (up to 0.12). The nature
of the numerical code (which does not include explicit diffusion) of
\cite{hawley99} did not allow them to make statements on critical
Reynolds numbers for the onset of turbulence in these flows, but the
results of \cite{lesur05} show good agreement with the experiments of
\cite{tillmark96} on the cyclonic side and seem to indicate that
transition occurs on the anticyclonic side for $\rey\sim 1500$ at $\rom=-1$. 
Also, their results show that the critical Reynolds number for the
appearance of turbulence at more negative rotation numbers increases
very rapidly ($\rey_c\sim 4\times 10^4$ for $\rom=-1.03$), which is not
consistent with the experimental results of  \cite{richard01}.
 \cite{ji06} have recently performed a new Taylor-Couette
   experiment. Their results show that quasi-Keplerian flows remain
   essentially  steady at Reynolds numbers up to millions, in agreement
   with the conclusions of \cite{lesur05}.



This summary of our current understanding of transition in non-rotating
and rotating shear flows makes it clear that we still do not apprehend
the whole process in a satisfying way. Some important questions, such as
the existence of a some qualitative symmetry with respect to $\rom=-0.5$
or the existence of a self-sustaining process in the presence of
rotation, remain largely unanswered. From our point of view, these
issues have to be addressed theoretically before conclusions on hydrodynamic
transition to turbulence in Rayleigh-stable rotating shear flows such as
Keplerian flows can be made. 



\section{Setting up the problem\label{equations}}
\subsection{Physical model}
In order to address this problem, we considered an
incompressible spanwise-rotating plane Couette flow between rigid 
boundaries, which is representative of shear flows that are linearly stable for
all Reynolds numbers in the absence of rotation. As is common in the
fluid dynamics literature, $(x,y,z)$ stand respectively for the streamwise
(azimuthal, $-y$ in the shearing sheet coordinates),  shearwise (radial,
$x$ in the shearing sheet coordinates) and spanwise (vertical)
directions.  The shearwise direction is bounded by two rigid walls, so
that $-d<y<d$, and $d$ is chosen as unit of length. We further assume
that the $x$ and $z$ direction are periodic with spatial periods 
$L_x=2\pi d/\alpha$ and $L_z=2\pi d/\beta$ . The background flow reads
$\vec{U}_B=S y\,\vec{e}_x$, the rotation vector is given by
$\vec{\Omega}=\Omega\,\vec{e}_z$ and
the constant shear rate $S$ is used to define a time unit.
In nondimensional form we therefore have $-1<y<1$ and
$\vec{U}_B=y\,\vec{e}_x$. The classical definition of the Reynolds
number in plane Couette flow is based on the shear velocity at the
boundaries and half of the channel width:
\begin{equation}
  \label{eq:reynolds}
  \rey=\f{|S| d^2}{\nu},
\end{equation}
where $\nu$ is the constant kinematic viscosity of the fluid. 
The Reynolds number defined above is four times smaller than
the ones used by \cite{nagata86} and \cite{lesur05}, and
the rotation number, defined in Eq.~(\ref{eq:rotation}), obeys
$\rom=-\Omega_{\mbox{\tiny TC}}/\rey_{\mbox{\tiny TC}}$ where the TC
index refers to quantities defined in the Taylor-Couette problem
by \cite{nagata86}.

\subsection{Equations} 
The full velocity field, denoted by $\vec{U}$, is split into
$\vec{U}_B$, the original streamwise background flow, and velocity
perturbations $\vec{u}=(u,v,w)$, which include any possible modification
of the original background flow.
The nondimensional equations governing the evolution of $\vec{U}$ and of
the total pressure $P=P_B+p$ (implicitly divided by the constant density
$\rho$), where $p$ denotes the pressure perturbation, are the
incompressible Navier-Stokes equations, which read
\begin{eqnarray}
\hspace{0.7cm}\difft{\vec{U}}+\vec{U}\cdot\vec{\nabla}\vec{U}-\rom\,\vec{e}_z\times\vec{U}
& =  & -\grad{P}+\frac{1}{Re}\Delta\vec{U},\\
\div{\vec{U}} & = & 0.
\end{eqnarray}
These equations have to be completed by periodic boundary conditions in $x$ and $z$
and by rigid (no-slip) boundary conditions in the $y$ direction
(the $x,z$ dependence is omitted here):
\begin{equation}
\label{eq:bc}
\displaystyle{u(-1)=u(1)=v(-1)=v(1)=w(-1)=w(1)=0,}
\end{equation}

\noindent Anticipating the numerics, an important point 
is that these equations reduce to two three-dimen\-sional scalar equations
governing the evolution of shearwise velocity $(v)$ and vorticity
$(\eta)$ perturbations \citep{schmid00}:
\begin{equation}
\label{eq:v}
 \left(\difft{}-\f{1}{\rey}\Delta\right)\Delta
 v+\vec{P}_v\cdot\left(\vec{U}\cdot\grad{\vec{U}}-\rom\,\vec{e}_z\times\vec{U}\right)=0,
\end{equation}
\begin{equation}
\label{eq:eta}
   \left(\difft{}-\f{1}{\rey}\Delta\right)\eta+\vec{P}_\eta\cdot\left(
\vec{U}\cdot\grad{\vec{U}}-\rom\,\vec{e}_z\times\vec{U}\right)=0,
\end{equation}
where
\begin{eqnarray}
  \label{eq:PvPeta}
\hspace{3.cm}  \vec{P}_v & = & -\vec{e}_y\cdot\curl{\curl{(\cdot)}},\\
  \vec{P}_\eta & = & \vec{e}_y\cdot\curl{(\cdot)},
\end{eqnarray}
plus two streamwise and spanwise-independent equations describing
the evolution of the mean streamwise and spanwise velocity fields:
\begin{equation}
  \label{eq:meanflow1}
  \left(\difft{}-\f{1}{\rey}\ddiffyy{}\right)\left<{u}\right>+
\left<{\vec{e}_x\cdot\vec{U}\cdot\grad{\vec{U}}}\right>=0,
\end{equation}
\begin{equation}
  \label{eq:meanflow2}
  \left(\difft{}-\f{1}{\rey}\ddiffyy{}\right)\left<{w}\right>+
\left<{\vec{e}_z\cdot\vec{U}\cdot\grad{\vec{U}}}\right>=0,
\end{equation}
where brackets denote $(x,z)$ averages. Note that $\left<v\right>=0$
results from the incompressibility constraint and boundary
conditions~(\ref{eq:bc}), and that the Coriolis force does not enter
Eqs.~(\ref{eq:meanflow1})-(\ref{eq:meanflow2}).

\subsection{Symmetries}
The fluid motions that will be studied here have several
symmetries, which can be used to reduce the computational costs. 
We will first deal with three-component two-dimensional (streamwise
independent) flows (hereinafter 2D-3C), like combined streamwise rolls (vortices)
and streaks, which have a reflection symmetry with respect to the $z=0$
plane for a properly chosen phase in $z$. This symmetry reads
\begin{equation}
  \label{eq:reflect}
  \mathcal{Z}: (x,y,z)\rightarrow (x,y,-z),\quad (v,\eta)\rightarrow (v,-\eta).
\end{equation}
Thanks to the $\mathcal{Z}$ symmetry of the basic 2D-3C flows, fully
three-dimensional nonlinear solutions bifurcating from these flows have
either the $\mathcal{Z}$ symmetry or the so-called ``shift-and-reflect"
symmetry given by
\begin{equation}
  \label{eq:shiftreflect}
  \mathcal{S}_2: (x,y,z)\rightarrow (x+L_x/2,y,-z),\quad
  (v,\eta)\rightarrow (v,-\eta).
\end{equation}
A basic 2D-3C solution has both $\mathcal{Z}$ and
$\mathcal{S}_2$ symmetries but the three-dimensional part of
the total velocity field has only one of them. Finally, choosing the
$x$-phase of the three-dimensional solution (brackets denote $(x,y)$
averages) such that 
\begin{equation}
  \label{eq:phasecond}
  \Im\left<\eta\exp\,(-i\alpha x)\right>=0\quad \mbox{or}\quad \Im\left<v\exp\,(-i\alpha x)\right>=0,
\end{equation}
depending on  whether the solution has $\mathcal{S}_2$ or
$\mathcal{Z}$ symmetry, it is possible to show that for plane Couette flow
the velocity field has the following ``shift-and-rotate'' symmetry:
\begin{equation}
  \label{eq:shiftrotation}
  \mathcal{S}_1:(x,y,z)\rightarrow (L_x/2-x,-y,z+L_z/2),\quad
  (v,\eta)\rightarrow (-v,\eta).
\end{equation}
The forementioned symmetries have been named according to the
nomenclature used in \cite{wedin04} with $(x,y,z)$ here corresponding to
$(z,r,\phi)$ in their paper on pipe Poiseuille flow (not to be confused
with the cylindrical coordinates sometimes used for accretion disks).

\subsection{Comments on the model}
Before we start describing our results, let us briefly comment on our
choices regarding this physical model. We will discuss
two points: incompressibility and boundary conditions. 
Accretion disks are fully compressible and our assumption of
incompressibility may appear as a limitation. We fully
support the idea that sound waves may play an important role in the
nonlinear dynamics of accretion disks, notably by destabilizing coherent
vortical structures \citep{broadbent79,bodo05} and by limiting their extent
because of the need to maintain causal acoustic contact, and that
compressible physics should be included in any accretion disk study
aiming at being realistic. However, the fact that disks are compressible
does not mean that incompressible dynamics is unimportant in this environment.
Our motivation in this study has been to identify such fundamental
incompressible processes, which are major actors of the
transition to turbulence in non-rotating shear flows. The
self-sustaining process described in Sect.~\ref{review}
does not require any compressible effect and is already fairly complex,
so our choice represents a natural simplification to tackle the
corresponding rotating problem.

The second  point has to do with our choice of a wall-bounded flow
model, which may not be relevant to the physics of accretion disks,
notably because walls are responsible for the appearance of boundary
layers and global modes. There are two reasons for this choice here: 
first, investigating  exact nonlinear dynamics using continuation
methods usually requires that solutions be steady in a
given frame of reference, and finite shearing boxes do not admit 
steady non-axisymmetric ($x$-dependent) solutions 
(non-axisymmetry is mandatory to have sustained hydrodynamic turbulent
angular momentum transport in Rayleigh-stable and Rayleigh-neutral
flows). In fact, using our set of coordinates and dimensional
units, the shearing-periodic boundary condition on $\bu(x,y,z,t)$ is
\begin{equation}
  \bu(x,y+L_y,z,t)=\bu(x-SL_yt,y,z,t).
\label{shearing-periodic}
\end{equation}
If $\bu(x,y,z,t)$ depends on $x$ (its first argument) but not on $t$
(its fourth argument) then Eq.~(\ref{shearing-periodic}) cannot
be satisfied because the r.h.s. depends on time, while the
l.h.s. does not. For this reason, the finite shearing box
does not offer a practical framework to investigate the exact nonlinear
dynamics of shear flows. It may admit non-axisymmetric
solutions that are periodic in time, but it seems technically difficult
to obtain them  using standard  numerical methods.
The second reason has to do with the phenomenology of rotating shear flow
turbulence in laboratory. At first glance, the fact that only 
wall-bounded flows admit steady three-dimensional nonlinear solutions
may be problematic for self-sustaining process theories in accretion
disks. However, we note that the phenomenology of the onset of
turbulence looks very similar in unbounded and wall-bounded flows:
for instance, transition Reynolds numbers observed for cyclonic
  rotation in shearing boxes or in configurations with rigid
  boundaries are comparable \citep{lesur05}. Also, experiments on
wall-bounded Rayleigh-stable anticyclonic plane shear flows face
exactly the same difficulty as shearing box simulations,
\textit{i.~e.} turbulence dies away for $\rom$ close to
-1 \citep{tillmark96,bech97} at Reynolds numbers $\mathcal{O}(10^3)$ for
both types of set-ups.
Studying the analytically and numerically simpler case of wall-bounded
flows may thus prove useful to understand the unbounded problem too.

\section{Numerical methods\label{nummethods}}
Bearing in mind the symmetry of Couette flow  with respect to the centre of the
channel, we aimed at computing \textit{steady} nonlinear solutions of the previous set of
equations (in plane Poiseuille flow, which does not have this symmetry,
solutions would be travelling). For this purpose, we
developed two codes: a full nonlinear continuation code similar to those
of \cite{waleffe03} and \cite{wedin04} to compute exact
three-dimensional nonlinear steady solutions of
Eqs.~(\ref{eq:bc})-(\ref{eq:v})-(\ref{eq:eta})-(\ref{eq:meanflow1})-(\ref{eq:meanflow2})
thanks to a Newton iteration algorithm,  and a two-dimensional linear stability
code to find accurate initial guesses for the nonlinear code.
In both codes, discretization is done on a Gauss-Lobatto  grid
(Chebyshev collocation) in the shearwise
direction and a Fourier spectral representation is used in the $x$ and $z$
directions. Boundary conditions are implemented using differentiation
matrices computed with the DMSuite package \citep{weidemann00}.
In the continuation code, nonlinear terms are computed in
physical space and the steady nonlinear equations are solved in
spectral-physical-spectral space  using a direct LAPACK
solver to perform Newton iterations. The stability code makes uses of 
either the direct LAPACK eigenvalue solver, which computes the whole spectrum
of eigenvalues of the discretized operator, or the iterative Arnoldi
ARPACK solver to compute individual eigenmodes.
The nonlinear code was first checked with a two-dimensional
subcritical test problem, namely the Orr-Sommerfeld problem for a
no-slip plane Poiseuille flow \citep{zahn74,herbert76} and the full
three-dimensional version was successfully tested by reproducing
the bifurcation diagram for no-slip non-rotating plane Couette flow 
obtained by \cite{waleffe03} for $(\alpha,\beta)=(0.577,1.15)$ (see below).
Dealiasing was implemented but did not prove necessary to obtain accurate
solutions in most cases.
 The linear solver was tested by calculating the limits and critical
wavenumbers for the centrifugal instability and by  quantitatively 
reproducing results obtained for the inflectional instability of streaks
\citep{waleffe95,waleffe97,reddy98} in both plane Couette flow and
plane Poiseuille flow, for which the instability of streaks takes the form
of a travelling wave (non-zero eigenfrequency). Pseudo-arclength
continuation was implemented to follow branches of solutions.
The previously mentioned symmetries can be turned on or off in the
nonlinear code in order to reach higher resolutions. We checked that the
introduction of the symmetries module of the code gave the same results
as the full problem for mild resolutions.  We would finally like to
emphasize here that the numerical methods used in our calculations for
spatial differentiation have spectral (exponential) convergence, and
that the maximum affordable resolutions $(N_x,N_y,N_z)=(32,24,32)$ used
in this paper (which require approximately one Gigabyte of memory) 
allow for the computation of fairly complex velocity fields
with an excellent precision.

\section{Cyclonic rotation\label{numerics_cycl}}
\subsection{Computation of exact nonlinear solutions in non-rotating plane Couette flow}
In this section, we report results obtained for cyclonic rotation. In
order to reach this regime, we first computed steady non-rotating plane
Couette flow solutions using the same technique as \cite{waleffe03} (see also
\cite{wedin04} for a similar approach for the pipe Poiseuille flow
problem). The maximum resolution used for these calculations was
$(N_x,N_y,N_z)=(32,24,32)$, which is similar to the
resolution of \cite{waleffe03}, and the $\mathcal{S}_1$ and
$\mathcal{S}_2$ symmetries were used in the code  to compute fully
three-dimensional solutions. 

The basic idea  of this forcing approach, as we will call it in
the remainder of the paper, is to compute fully three-dimensional
nonlinear solutions of the forced Navier-Stokes equations using Newton
iteration, then decreasing progressively the forcing term until a
solution for zero forcing is found. 
In practice, an initial guess is required to perform this calculation.
In this study, we chose $(\alpha,\beta)=(0.577,1.15)$, for which the
global critical Reynolds number for the self-sustaining process is known
to be minimized, and  set $\rey=150$. The determination of the initial
guess was done in two steps. The first step was to compute the 2D-3C
streamwise-independent part of the flow, denoted by
$(u_\twod,v_\twod,w_\twod)$ and the second step was to compute the
three-dimensional instability of this 2D-3C flow. We first focus on the
first step, which relies on the physics of the lift-up effect in order to
calculate a combined rolls and streaks velocity field.
The simplest streamwise rolls structure is given by the most slowly decaying
solution of the linear Stokes problem for the streamwise vorticity
$\omega=\exp\,(\lambda t/\rey)\,\omega_s(y,z)$ 
\begin{equation}
  \label{eq:stokes}
\lambda\,\omega_s=\Delta\,\omega_s,
\end{equation}
where $\lambda$ is negative. For rigid no-slip boundary conditions and
a $z$-periodicity of the rolls $\beta=1.15$, the least stable eigenvalue
is given by $\lambda=-9.27306$.  The corresponding Stokes velocity field is
\begin{equation}
(v_s,w_s)=(-\partial_z\Delta^{-1}\omega_s,\partial_y\Delta^{-1}\omega_s).
\end{equation}
The physics of the lift-up effect is such that $\mathcal{O}(1/\rey)$
streamwise rolls generate $\mathcal{O}(1)$ streamwise velocity, so we
chose the amplitude of the rolls to be
\begin{equation}
(v_\twod,w_\twod)=\f{A}{\rey}\f{(v_s,w_s)}{\mbox{max}(v_s)},
\end{equation}
where $A$ is assumed to be $\mathcal{O}(1)$. As shown by the sign of
$\lambda$, this rolls field decays viscously, so that the following
$\mathcal{O}(1/\rey^2)$ forcing term $\vec{F}$ 
\begin{equation}
  \label{eq:forcing_stokes}
  (F_x,F_y,F_z)=\f{|\lambda|}{\rey} (0,v_\twod,w_\twod)=\f{A|\lambda|}{\rey^2}\f{(v_s,w_s)}{\mbox{max}(v_s)}
\end{equation}
had to be included in the Navier-Stokes equations in order to compensate
for viscous diffusion. The determination of the 2D-3C part of the
initial guess was finally  completed by solving the linear problem 
\begin{equation}
  \label{eq:lin_streaks}
  \left(\f{1}{\rey}\Delta - v_\twod\diffy{} -w_\twod\diffz{}\right) u_\twod= v_\twod \diffy{U_B}
\end{equation}
\noindent
for the $\mathcal{O}(1)$ streaks field $u_\twod$. For a
well-chosen $z$ phase, this 2D-3C field has $\mathcal{Z}$ and
$\mathcal{S}_2$ symmetries.

\begin{figure}[ht]
\resizebox{\hsize}{!}{%
\includegraphics[width=10cm]{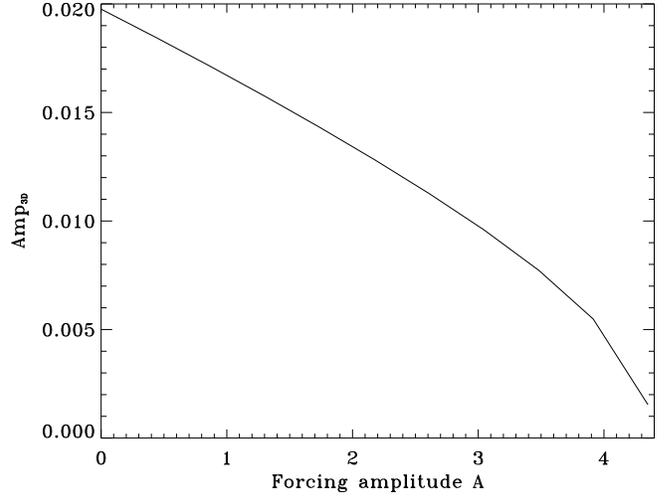}}
   \caption{Bifurcation diagram of forced nonlinear solutions in
     non-rotating plane Couette flow with
     $(\alpha,\beta)=(0.577,1.15)$ with respect to the forcing amplitude
     $A$, for $\rey=150$. $\mbox{Amp}_\threed$, defined by
     Eq.~(\ref{eq:defamp}), is used to measure the amount of
     three-dimensionality in the full nonlinear solution. The resolution
     $(N_x,N_y,N_z)=(32,24,32)$ was used to perform this calculation.}
   \label{fig:force_bif}
 \end{figure}

The three-dimensional part of the initial guess was then determined 
by computing a neutral instability mode of the rolls and
streaks field for the chosen value of $\alpha$ using the linear
stability solver. For $\alpha=0.577$, this mode, which has
$\mathcal{S}_2$ symmetry, is marginally stable when the forcing
amplitude is $A\simeq 4.4$.  Using the total rolls and streaks field and a small
three-dimensional neutral mode perturbation as an initial guess, we  finally
succeeded in computing an exact three-dimensional nonlinear steady
forced solution using the Newton solver.
In order to obtain a solution for zero forcing, we then only had 
to reduce the amplitude of the forcing term step-by-step down to zero
amplitude and to follow the corresponding forced solution using the
nonlinear continuation code. 
The bifurcation diagram with respect to the forcing amplitude $A$ is
shown in Fig.~\ref{fig:force_bif}. In this plot, the $y$-averaged amplitude of 
the $k_x=\alpha$ mode of the shearwise vorticity $\eta$
\begin{equation}
\label{eq:defamp}
\mbox{Amp}_\threed=\Re\int_{-1}^{1}\!\!\!\left<\eta\exp\,(-i\alpha x)\right>\mbox{d}y
\end{equation}
is used to measure the amount of three-dimensionality of the solution.
As shown in this diagram, the whole procedure works because the
bifurcation to three-dimensional solutions is subcritical with respect
to the forcing amplitude. Note also that $A$ is effectively
$\mathcal{O}(1)$ at the bifurcation point.

\begin{figure}[ht]
\resizebox{\hsize}{!}{%
\includegraphics[width=10cm]{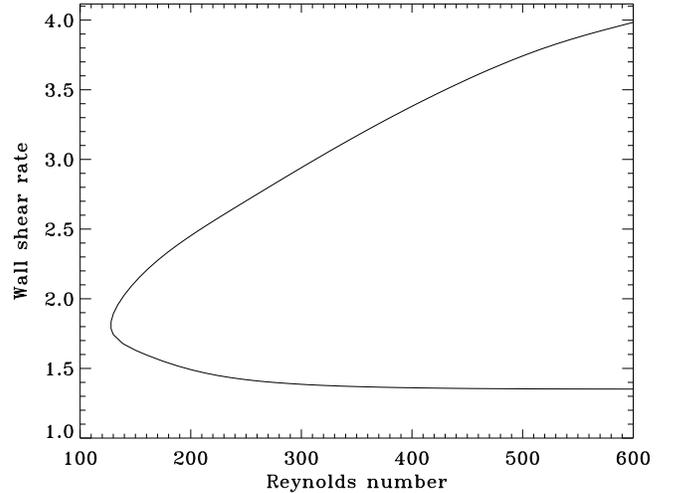}}
   \caption{Bifurcation diagram for non-rotating plane Couette flow with
 $(\alpha,\beta)=(0.577,1.15)$ with respect to $\rey$. The
 nondimensional wall shear rate $|\partial
     \left<{U}\right>/\partial y|(y=\pm 1)$ is used as a measure of
 nonlinearity (for the laminar flow, $|\partial \left<{U}\right>/\partial
 y|(\pm 1)=1$).}
   \label{fig:bifcouette}
 \end{figure}

Before cyclonic continuation was attempted, the non-rotating plane
Couette flow solution obtained previously was finally continued with
respect to the Reynolds number in order to obtain the
full bifurcation diagram for the chosen aspect ratio. The Reynolds
number of the saddle-node bifurcation point is $\rey=127.7$ (see
Fig.~\ref{fig:bifcouette}). 

\subsection{Continuation in the cyclonic regime}
\begin{figure}[ht]
\resizebox{\hsize}{!}{%
\includegraphics[width=10cm]{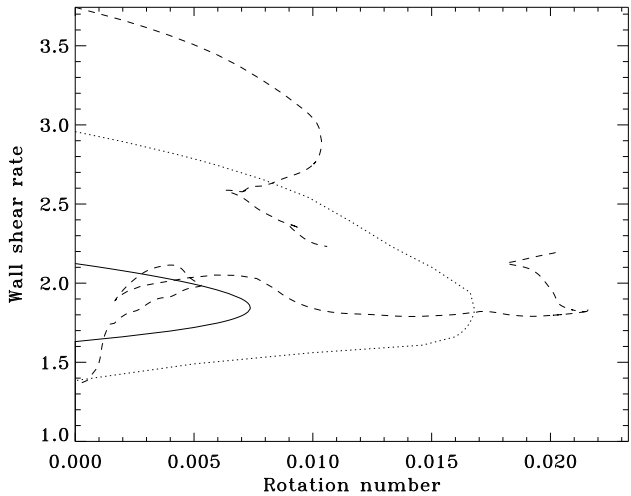}}
   \caption{Branches of three-dimensional nonlinear solutions obtained
     in the cyclonic region for $\rey=150$ (full line), $\rey=303$
     (dotted line) and $\rey=500$ (dashed line). A turning point 
     corresponding to a critical rotation number is found for any
     $\rey$. The $\rey=500$ calculation suffered from a lack of numerical
     resolution.}
   \label{fig:branchescyclo}
 \end{figure}

\begin{figure}[ht]
\resizebox{\hsize}{!}{%
\hspace{0.2cm}\includegraphics[width=9cm]{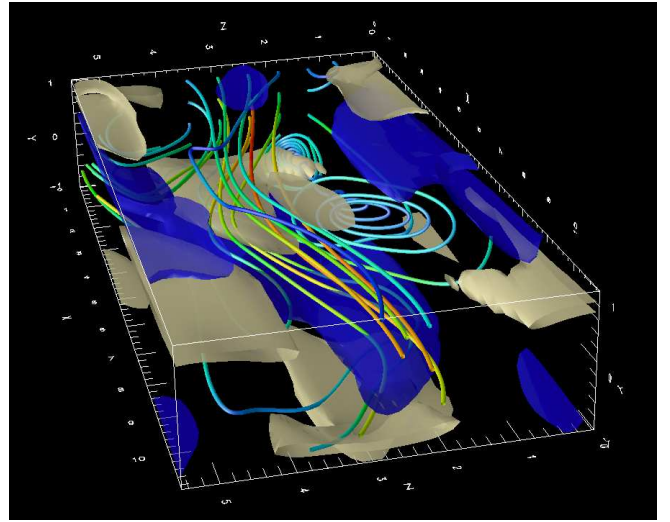}\hspace{0.2cm}}
   \caption{Positive (light brown) and negative (dark blue)
     isovorticity surfaces and randomly selected streamlines  for
     the nonlinear solution located at $\rey=303$ and $\rom=0.017$.}
   \label{fig:snapcyclo}
 \end{figure}

\begin{figure}[ht]
\resizebox{\hsize}{!}{%
 \includegraphics[width=10cm]{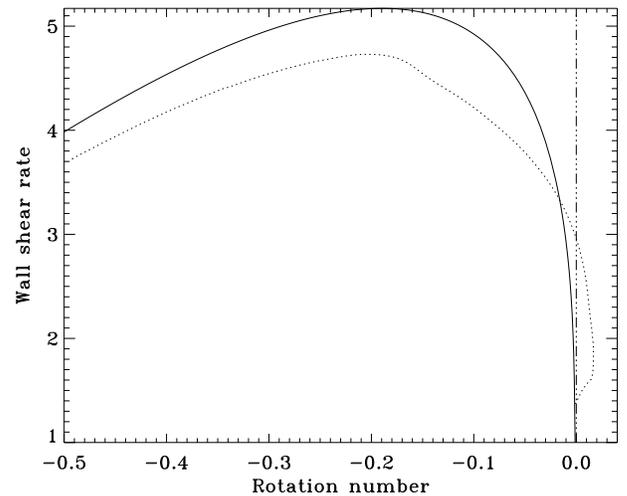}}
   \caption{Bifurcation diagram for Taylor vortices (full line) showing
           their connection with three-dimensional rotating
           self-sustaining solutions (dotted line) at $\rey=303$. 
           The dash-triple dotted curve shows the centrifugal stability
           limit. $(\alpha,\beta)=(0.577,1.15)$.}
   \label{fig:nagplot}
 \end{figure}

The bifurcation diagram obtained in the previous paragraph was then used
to compute solutions in the $\rom>0$ region by performing continuation with
respect to $\rom$. Solutions for several Reynolds numbers were continued
from the lower branch of Fig.~\ref{fig:bifcouette}. 
A turning point is found systematically at a critical
rotation number depending on the chosen Reynolds number. 
The solution then evolves back to zero rotation number
and reaches the upper branch  of the non-rotating plane Couette flow
bifurcation diagram. For the forementioned maximum affordable
resolution, solutions could be obtained for Reynolds
numbers up to 500. Note that  we suffered from a lack of numerical
resolution for $\rey=500$. The  continuation curve is incomplete in that case 
and is given for illustrative purposes only, because the
corresponding results are probably not very accurate. Unfortunately, it is not possible to
increase the resolution of this experiment further by a significant amount
because of both memory and CPU requirements. For instance, the LAPACK
linear solver has algorithmic complexity $\mathcal{O}(N^3)$, where $N$
is the system size, so that doubling the resolution in even only one
direction would be prohibitive (computing one point of the continuation
curves using the present resolution already takes several hours on a 3
GHz processor). Branches of solutions obtained at various Reynolds
numbers are plotted in Fig.~\ref{fig:branchescyclo} and a 
snapshot of isovorticity surfaces and streamlines for the solution  at the 
turning point at $\rey=303$ and $\rom=0.017$ is shown in
Fig.~\ref{fig:snapcyclo}. Anticipating a comparison with results for
anticyclonic rotation, we also present in Fig.~\ref{fig:nagplot} a
bifurcation diagram for $\rey=303$ similar to the one obtained by
\cite{nagata90}, showing both the 2D-3C Taylor vortices branch in the
linearly unstable weakly anticyclonic regime and its connection with the
three-dimensional wavy Taylor vortices penetrating into the cyclonic
region. 

\begin{figure}[ht]
\resizebox{\hsize}{!}{%
 \includegraphics[width=10cm]{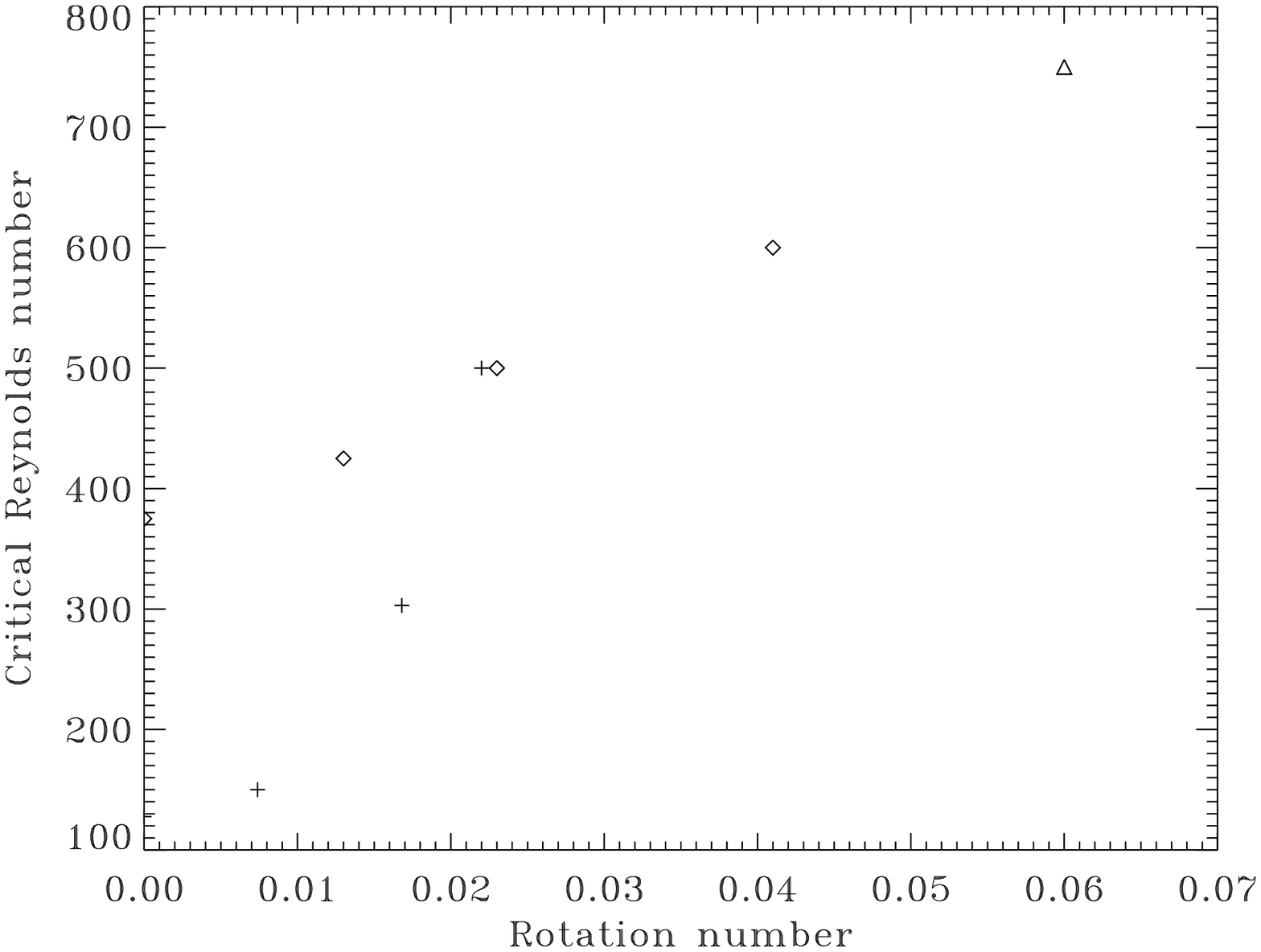}}
   \caption{Critical Reynolds number for   exact three-dimensional nonlinear solutions for
     $(\alpha,\beta)=(0.557,1.15)$ as a function of the
     rotation number $\rom$ (+),  compared to the  experimental data obtained
      by \cite{tillmark96} ($\diamond$) for the onset of turbulence and the
     numerical data  by \cite{komminaho96}  ($\triangle$).  
     The $\rey=500$ point only gives a qualitative indication of the
     actual critical value because of the lack of 
     resolution for the corresponding numerical experiment.
     The numerical results of  \cite{lesur05} for shearing box
     simulations match those obtained by \cite{tillmark96}  and are not
     reproduced here.}
   \label{fig:Re-Ro-cyclo}
 \end{figure}

It is interesting to compare the critical $\rom-\rey$ curve obtained
from our computations with already published experimental and numerical work.
Figure~\ref{fig:Re-Ro-cyclo} shows that the critical $\rey$
obtained for the exact coherent structures is smaller
than the critical $\rey$ for the onset of turbulence in cyclonic
experiments for $\rey=150$ and $\rey=303$ (note again that the
$\rey=500$ point in that plot is purely illustrative due to the lack of
numerical resolution). The critical $\rey$ is seen to increase with
increasing cyclonic rotation, similarly
to experiments. A possible interpretation for these observations is that
the  values obtained in our study for $(\alpha,\beta)$ parameters that
minimize the global critical Reynolds number for the non-rotating plane
Couette flow  provide lower bounds on the transition $\rey$  in the
cyclonic regime as well. Since \cite{tillmark96} and \cite{lesur05} used
a different aspect ratio in their experiments, 
larger Reynolds numbers may be required in their set-up to obtain
turbulence. An alternative explanation is that the exact
coherent structures computed here do not allow for a precise
determination of the turbulence threshold in a real experiment 
and only give a rough estimate for the critical $\rey$, because the
actual transition process involves not only these solutions but 
all their possible instabilities, as mentioned previously.

\section{Anticyclonic rotation\label{numerics_anti}}
Several arguments regarding the existence of turbulence
in Keplerian flows have been based on a possible symmetry with respect
to $\rom=-0.5$. In this section, we report some continuation experiments 
in the anticyclonic region, which notably show that the self-sustaining
process present in non-rotating or weaky cyclonic shear flows
($\rom\simeq 0$) cannot be transposed into the 
$\rom\simeq-1$ regimes and that more complex nonlinear structures have
to be searched for to find subcritical transition  (if any) in this
region. 
\subsection{Finding a starting point for continuation}
Nonlinear continuation methods require a starting point, which
is usually a bifurcation point. This bifurcation can have a
physical origin, when a linear instability like the centrifugal
instability is present. In that case, continuation is
straightforward. When there are no linear instabilities, a possibility
is to include an extra artificial term in the original equations to
trigger a linear instability that would otherwise be absent, and to
study the subsequent nonlinear dynamics when this artificial term is
gradually removed. This is how \cite{nagata90} proceeded in order to
find three-dimensional solutions of non-rotating plane
Couette flow. By adding a small amount of anticyclonic rotation to the original
Couette flow, he caused the background flow to become centrifugally unstable and
followed the 2D-3C Taylor vortices (axisymmetric structures with small
amplitude streamwise rolls and strong streamwise velocity) into the
nonlinear domain. In this  regime, Taylor
vortices become unstable to three-dimensional wavy Taylor vortices at a
critical Reynolds number depending on the amount of rotation of the
system. The interesting finding of \cite{nagata90} was that he was able
to follow this three-dimensional solution down to zero rotation, thanks
to a proper nonlinear feedback of the three-dimensional structures on the 
streamwise rolls  originally generated by the centrifugal instability. In
the previous section, we  proceeded in a similar way
to find non-rotating solutions, but instead of adding a small amount of
anticyclonic rotation originally, we used the same method as
\cite{waleffe03} to sustain the rolls field, by adding an artificial forcing
term in the streamwise vorticity equation. A third possibility, found by
\cite{clever92}, is to cause the Couette flow to become unstable to
convection by imposing a supercritical
temperature gradient in the shearwise direction in order to force the
rolls field, and then perform a nonlinear continuation of the secondary
instabilities of the convection rolls when the temperature gradient is
suppressed. In spite of their \textit{a priori} rather different characteristics,
all these approaches lead to the same result. The reason for this is
linear transient growth \textit{via} the lift-up effect. All these methods
sustain weak streamwise rolls which give rise to a finite amplitude streaks velocity
field, as a consequence of the spanwise redistribution of streamwise velocity. The
instability of streaks, whatever the origin of streamwise rolls is,
results from the presence of inflection points in the spanwise streak
profile \citep{jones81}.

In order to investigate the nonlinear dynamics in the rotating Keplerian
regime, a proper starting point is required too, and a major problem
is to find it. Unfortunately, we did not find a nonlinear connection between
the non-rotating plane Couette flow solutions and the strongly
anticyclonic region. As shown in Fig.~\ref{fig:nagplot}, the
three-dimensional lower branch of solutions that crosses the $\rom=0$ line
joins the two-dimensional Taylor vortex branch at some very small negative
$\rom$, and we had difficulties in following the three-dimensional upper
branch for $\rom< -0.5$ for the parameters of
Fig.~\ref{fig:nagplot} (the solutions on this branch turn into
nonlinear travelling waves  (non-zero phase speed) at
$\rom\simeq-0.216$, similarly to what has been observed in experiments
\citep{andereck86,nagata90}). Thus, another starting point must be found
in order to study the region of large
negative $\rom$. It seemed difficult to start directly from the
Keplerian regime, which is very far from any linear instability
point, so our idea was to start from more convenient rotating regimes
and to try to continue possible solutions towards $\rom=-4/3$. We
experimented with two ideas for $\rom=-1$ or close to minus one.

The first working hypothesis, inspired by \cite{nagata90}, was to start
from values of $\rom$ larger than but close to -1 in order to trigger a
centrifugal instability, and to study the nonlinear behaviour of
secondary instabilities of the Taylor vortices in that regime, when
$\rom$ is made closer to -1 (centrifugal instability approach). 
The second technique, similar to that used in Sect.~\ref{numerics_cycl}
to obtain solutions for $\rom=0$, was motivated by a very strong analogy and
near-symmetry regarding \textit{linear theory} between the $\rom=0$ and $\rom=-1$
regimes. From the strictly linear point of view, \cite{antko06} have shown that 
the lift-up effect, which occurs for $\rom=0$, has an equivalent for
$\rom=-1$, christened the ``anti lift-up'' effect. This equivalence can
be seen by considering the linear equations describing the
evolution of the streamwise velocity field $u$ and of the streamwise
vorticity $\omega$ (the rolls field) in plane Couette flow with system
rotation:
\begin{eqnarray}
  \label{eq:streaks}
  \difft{u} & = &  -(\rom+1)v+\f{1}{\rey}\Delta u,\\
  \label{eq:rolls}
\hspace{2cm}  \difft{\omega} & = & -\rom \diffz{u} +\f{1}{\rey}\Delta\,\omega.
\end{eqnarray}

\noindent For $\rom=0$, a streamwise velocity field is generated by
redistribution of shear by the streamwise rolls velocity field, leading to
strong transient amplification and the formation of streaks. 
There is no source term in Eq.~(\ref{eq:rolls}) and any streamwise
vorticity perturbation must decay viscously in this linear framework.
For $\rom=-1$, strong streamwise vorticity can be generated
transiently by the Coriolis force acting on the streamwise velocity.
The streamwise velocity equation~(\ref{eq:streaks}) has no source term
in that case because the Coriolis force acting on the rolls exactly compensates the
redistribution of the mean shear by the rolls, so that  streamwise velocity
perturbations decay viscously. Transient growth observed for
$\rom=-1$ can also be interpreted as an epicyclic oscillation
of infinite period, since the epicyclic frequency is zero in
this regime \citep{antko06}. Pushing the analogy with the non-rotating
case further and now considering nonlinear effects, the generation of
finite-amplitude rolls by the anti lift-up effect may then allow for the
development of three-dimensional instabilities that could behave in a
similar way to their $\rom=0$ counterparts and feedback on a weak
streamwise velocity field, thus leading to a self-sustaining process for
$\rom=-1$. The procedure in that case is to set directly  $\rom=-1$, to
artificially force a weak streaks field in order to amplify
streamwise rolls, and to follow the nonlinear development of the
instabilities of the rolls when removing progressively the forcing term
(forcing approach). 

\subsection{Centrifugal instability approach}
We start by describing results obtained in the first working
hypothesis. Using both linear stability codes and nonlinear
continuation, we identified streamwise-independent Taylor vortices
bifurcating from the spanwise-rotating background Couette flow at
various $\rom$ larger than but very close to -1. In practice, the
bifurcation to Taylor vortices occurs when the Taylor number
\begin{equation}
\label{eq:taylor}
\tay=-16\rey^2\rom (1+\rom)
\end{equation}
exceeds 1707, so that the critical $\rom$ depends only slightly on
$\rey$ at large $\rey$. We chose $\beta=1.5585$, which is the critical
wavenumber for this instability (in the Taylor-Couette experiment, this
would be $\beta=3.117$ because $2d$ is used as unit of length  in that case).
The structure of the full velocity field of the
Taylor vortices in this rotating regime is quite different from the one
at small $\rom$: close to $\rom=-1$, the streaks field has a lower
amplitude than for small $\rom$, which is a consequence of the
absence of the lift-up effect. The secondary instabilities of the Taylor
vortices close to the Rayleigh line $\rom=-1$ are therefore different
from those at small $\rom$, whose inflectional nature is due to the
shape and amplitude of the streamwise velocity field in that regime. This marks
a fundamental difference between the two regimes and already hints that
the argument on symmetries of solutions with respect to $\rom=-0.5$ is
limited. Due to the $\mathcal{Z}$ symmetry
of the Taylor vortices, two steady three-dimensional instabilities can
be found, with either $\mathcal{S}_2$ (wavy Taylor vortices, hereinafter
WTV) or $\mathcal{Z}$ (twisted Taylor vortices, hereinafter TTV)
symmetry. These instabilities have been identified by
\cite{nagata86} and TTVs have been observed experimentally by
\cite{andereck83} for $\rom\sim -0.8$. Finally, some steady subharmonic
instabilities with spanwise wavelength twice the wavelength of the
Taylor vortices and an instability with non-zero phase speed are also
present \citep[see][]{nagata88}. These latter instabilities are poor
candidates for continuation into the Rayleigh-stable regime because they
have been reported experimentally for $-0.5>\rom>-0.8$ only
\citep[see][]{andereck83}. Consequently, they have not been considered in
this study.

\begin{figure}[ht]
\resizebox{\hsize}{!}{%
 \includegraphics[width=10cm]{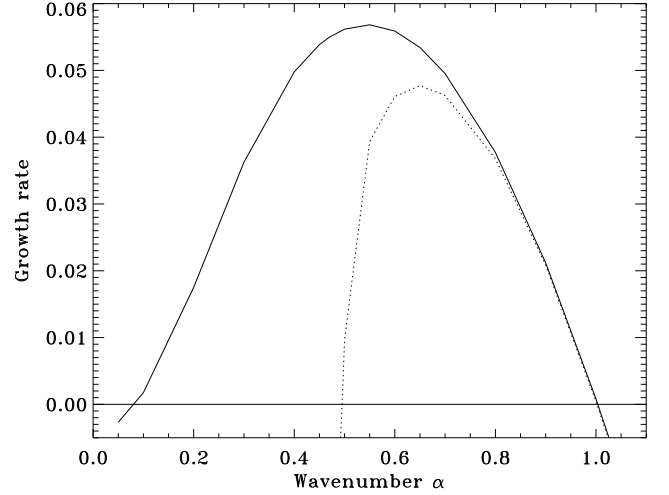}}
   \caption{Growth rate of the three-dimensional instability modes of
     the Taylor vortices as a function of the streamwise wavenumber
     $\alpha$  for $\rom=-0.97875$, $\rey=500$ and $\beta=1.5585$. Full
     line: wavy vortex mode. Dotted line: twisted vortex
     mode. Nonlinear continuation from the bifurcation point is
     initialized  by adding a small amount of neutral
     modes with  $\alpha=1$ to the Taylor vortex flow.}
   \label{fig:growthrate_centri}
 \end{figure}

\begin{figure}[ht]
\resizebox{\hsize}{!}{%
 \includegraphics[width=10cm]{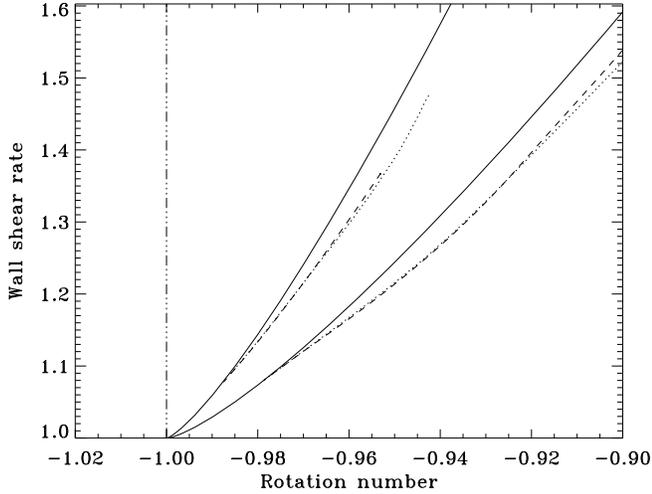}}
   \caption{Bifurcation diagrams in the $\rom\sim -1$ region for
     $(\alpha,\beta)=(1,1.5585)$ and Reynolds numbers $\rey=500$ (lower full,
      dotted and dashed lines) and $\rey=1500$ (upper full, dotted and dashed
     lines). The resolution is $(N_y,N_z)=(24,48)$ for
     the Taylor vortex branches  (full lines) and
     $(N_x,N_y,N_z)=(16,24,48)$ for the WTV branches (dotted
     lines) and TTV branches (dashed lines). The dashed-triple dotted
      line is the Rayleigh line. 
     Continuation was stopped at smaller values of
     $\rom$ for $\rey=1500$ due to insufficient numerical resolution.}
   \label{fig:bif_WTV}
 \end{figure}

Close to the Rayleigh line, the first unstable three-dimensional mode (when
$\rey$ is increased) is the wavy Taylor vortex mode for the numerically accessible
range of $\rey$ (see Fig.~\ref{fig:growthrate_centri} for a comparison
between the growth rate of WTVs and TTVs for some of the parameters used here).
In Fig.~\ref{fig:bif_WTV}, we show the corresponding bifurcation
diagrams in the vicinity of $\rom=-1$ for $\rey=500$ and
$\rey=1500$. The bifurcations to 2D-3C Taylor vortices are supercritical
with respect to $\rom$ and the associated branches of nonlinear
solutions do not cross the Rayleigh line $\rom=-1$. Independently of
$\rey$, the secondary bifurcations to three-dimensional WTVs are unfortunately also
supercritical and the corresponding branches of solutions do not approach
the centrifugal stability limits either, unlike their counterpart at
$\rom\sim 0$, in spite of having the same $\mathcal{S}_2$ symmetry
(compare with Fig.~\ref{fig:nagplot}).  The same remarks apply to TTVs.
This demonstrates clearly that the phenomenology of self-sustaining
processes at $\rom=0$ cannot be transposed directly into the $\rom=-1$
regime, even though analogous linear transient amplification mechanisms
exist in each of these regimes.

\subsection{Pitfalls of the forcing approach}
As mentioned earlier, using a forcing approach similar to that
presented in Sect.~\ref{numerics_cycl} appeared as a plausible
alternative to the centrifugal instability approach in order to identify
a self-sustaining process at $\rom=-1$. Such an attempt was further
encouraged by the work of  \cite{wedin04}, who managed
to identify a self-sustaining process in non-rotating pipe Poiseuille
flow by continuing forced nonlinear solutions, whereas \cite{barnes00}
had previously been unable to continue nonlinear solutions
obtained for a rotating pipe Poiseuille flow down to zero rotation.

The simplest (and possibly naive) way of tackling the problem was to
include an artificial forcing term 
\begin{equation}
  \label{eq:forcing}
  F_x(y,z)=\f{A}{\rey^2}\left(\beta^2+\f{\pi^2}{4}\right)\cos\beta
  z\cos\left(\f{\pi y}{2}\right)
\end{equation}
in the streamwise velocity equation in the Rayleigh-neutral $\rom=-1$
regime in order to sustain the following streamwise velocity field
against viscous dissipation:
\begin{equation}
\label{eq:streamvel}
  u(y,z)=\f{A}{\rey}\cos\beta z\cos\left(\f{\pi y}{2}\right).
\end{equation}
The forcing amplitude $A$ is assumed to be $\mathcal{O}(1)$, as in the
non-rotating problem  \citep[see][]{waleffe95,waleffe03}, in order 
to generate possibly unstable finite-amplitude $\mathcal{O}(1)$ streamwise
thanks to the Coriolis force acting on the $\mathcal{O}(1/\rey)$
streamwise velocity field. Note that unlike for
$\rom=0$, there is no need to solve numerically a Stokes
problem for the streamwise velocity here, because the boundary
conditions on $u$ make it possible to use exact expressions like
Eq.~(\ref{eq:forcing}) for the forcing term. 

\begin{figure}[ht]
\resizebox{\hsize}{!}{%
 \includegraphics[width=10cm]{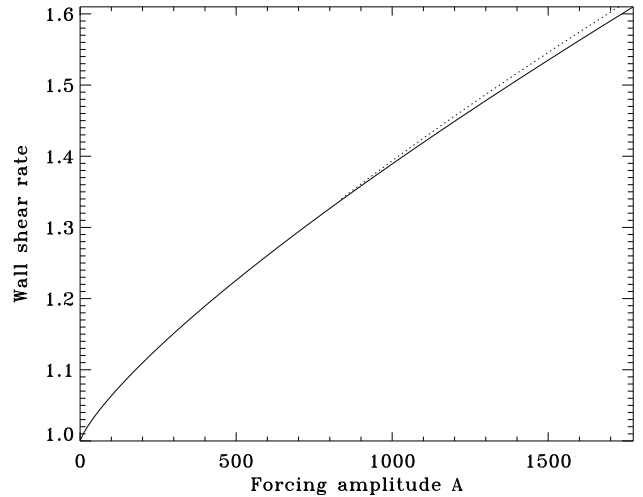}}
   \caption{Bifurcation diagram with respect to the forcing amplitude $A$
     for  $\rom= -1$, $(\alpha,\beta)=(0.8,1.5585)$ and $\rey=500$. The full
     line curve is the branch of 2D-3C forced solutions and the dotted
     line curve is the three-dimensional branch of solutions with $\mathcal{Z}$
     symmetry. The bifurcation to these three-dimensional solutions
     occurs for $A=834.7$   and is supercritical with respect to the
     forcing amplitude.}
   \label{fig:bif_force_anti}
 \end{figure}

Initializing the Newton solver with the streamwise velocity
field~(\ref{eq:streamvel}) and streamwise forcing
term~(\ref{eq:forcing}) plus a streamwise rolls field with streamwise
vorticity $\omega$ being the solution of the linear problem
\begin{equation}
  \label{eq:rolls_rom=-1} 
\Delta\,\omega=  -\rey\diffz{u},
\end{equation} 
nonlinear steady 2D-3C forced solutions could be obtained for
various forcing amplitudes (see Fig.~\ref{fig:bif_force_anti}).
However, their properties turned out to be quite
different from the prediction of linear theory that
the streamwise rolls velocity field be $\mathcal{O}(\rey)$ larger
than the streamwise velocity field. One of the reasons for this
difference is that the rolls feed back on the streamwise velocity field
which generates them \textit{via} the anti lift-up effect and sweep it
close to the boundaries.
In comparison, the corresponding effect at $\rom=0$ would be the
advection of axisymmetric structures like rolls by the streaks field
(the $u\,\partial_x$ operator), which is strictly zero due to the
streamwise-independence of the rolls and streaks pattern.
Therefore, the analogy between $\rom=0$ and $\rom=-1$ breaks down when
nonlinearity is included, at least for the forcing term used here (the
precise reasons for this break down and the choice of a suitable forcing term
will be further discussed in Sect.~\ref{theory}). The problem becomes
even more apparent when computing the stability of these 2D-3C solutions 
to three-dimensional infinitesimal perturbations with
wavenumber $\alpha$. For $\rey=500$ and $(\alpha,\beta)=(0.8,1.5585)$
for instance, bifurcation to three-dimensional solutions with
$\mathcal{Z}$ symmetry occurs for $A=\mathcal{O}(10^3)$ only, whereas
bifurcation occurs for $A=\mathcal{O}(1)$ in the $\rom=0$
case. Furthermore, the first three-dimensional instability
found for $\rom=-1$ is supercritical with respect to the forcing
amplitude (Fig.~\ref{fig:bif_force_anti}) and subsequent
three-dimensional nonlinear solutions cannot be tracked down to
zero forcing (compare with Fig.~\ref{fig:force_bif} for $\rom=0$).

This second numerical attempt confirms the conclusions drawn from the
centrifugal instability approach, namely that a self-sustaining process
close to the Rayleigh line, if any, must act quite differently from its
non-rotating counterpart. It could be argued that our values of $\rey$
are too small to observe
anything interesting in the anticyclonic regime, or that the chosen aspect ratio
does not allow for three-dimensional self-sustained
solutions. We performed the same kind of
computations with different aspect ratios for Reynolds numbers up to
approximately 1500 but did not notice any qualitative
change in the bifurcation diagrams, which always show bifurcation to
three-dimensional solutions at large $A$ and supercritical behaviour
with respect to the forcing amplitude.


\section{Asymptotic description of self-sustaining processes\label{theory}}
The  numerical results presented in the previous section illustrate
the difficulty of identifying self-sustaining processes in
Rayleigh-stable regimes. However, there is currently no mathematical
proof that such solutions do not exist, so that it worth attempting
to understand why they cannot be found in plane Couette flow on the
Rayleigh line and under which circumstances they could be found.
In this section, we provide some analytical arguments based on an
asymptotic description of the problem at large $\rey$ which, together
with the numerical results, highlight the fundamental differences
between the $\rom=0$ and $\rom=-1$ regimes in plane Couette
flow. Starting with the non-rotating case, we obtain a two-dimensional
formulation for the solutions of the lower branch of the non-rotating
self-sustaining process at very large $\rey$. We then consider several
possible asymptotic descriptions of three-dimensional solutions close to the
Rayleigh line and demonstrate why the actual Taylor vortices generated
in this regime in plane Couette flow cannot lead to a rotating
self-sustaining process, unlike their $-\rom\ll 1$ counterparts. 
We finally show that self-sustaining solutions on the Rayleigh-line 
may only exist if the 2D-3C part of the flow satisfies 
nontrivial solvability conditions.

\subsection{Non-rotating Couette flow}
\subsubsection{Basic equations}
Following Sect.~\ref{equations}, perturbations $\bu=(u,v,w)$ and $p$ to the
non-rotating plane Couette flow $\vec{U}_B=y\,\be_x$ satisfy the nonlinear
equation
\begin{equation}
\label{eq:nonrotating}
  (\p_t+y\p_x+\bu\cdot\bnabla)\,\bu+v\,\be_x=-\bnabla p+\epsilon\Delta\bu,
\end{equation}
together with suitable boundary conditions and the constraint
\begin{equation}
  \bnabla\cdot\bu=0.
\end{equation}
The parameter $\epsilon=1/\rey$ will be used in the following to perform
an asymptotic expansion.

\subsubsection{Streamwise averaging}
Let an overbar denote averaging in the streamwise
($x$) direction, for which periodic boundary conditions apply.  Then
$\bu$ and $p$ can be separated into ``mean'' and ``wave'' parts, \textit{i.~e.}
\begin{equation}
  \bu(x,y,z,t)=\bar\bu(y,z,t)+\bu'(x,y,z,t),
\end{equation}
\begin{equation}
  p(x,y,z,t)=\bar p(y,z,t)+p'(x,y,z,t),
\end{equation}
with $\overline{\bu'}={\bf0}$ and $\overline{p'}=0$.  We also separate $\bar\bu$ into
toroidal (streamwise) and poloidal (shearwise and spanwise) parts,
parallel and perpendicular to the averaging direction:
\begin{equation}
  \bar\bu=\bar u\,\be_x+\bar\bu_\rmp.
\end{equation}
The mean (toroidal and poloidal) and wave parts of Eq.~(\ref{eq:nonrotating}) are then
\begin{equation}
  (\p_t+\bar\bu_\rmp\cdot\bnabla)\,\bar u+\bar v=\epsilon\Delta\bar u+
  F_x,
\label{uxbar}
\end{equation}
\begin{equation}
  (\p_t+\bar\bu_\rmp\cdot\bnabla)\,\bar\bu_\rmp=-\bnabla\bar p+
  \epsilon\Delta\bar\bu_\rmp+\bF_\rmp,
\label{upbar}
\end{equation}
\begin{equation}
  (\p_t+y\p_x+\bar\bu\cdot\bnabla)\,\bu'+\bu'\cdot\bnabla\bar\bu+
  \bu'\cdot\bnabla\bu'+\bF+v'\,\be_x=-\bnabla p'+\epsilon\Delta\bu',
\label{u'}
\end{equation}
with
\begin{equation}
  \bnabla\cdot\bar\bu_\rmp=\bnabla\cdot\bu'=0,
\end{equation}
where
\begin{equation}
\label{eq:meanforce}
  \bF=-\overline{\bu'\cdot\bnabla\bu'}
\end{equation}
is the mean force associated with the wave.  We also define
\begin{equation}
  \bG=\bnabla\times\bF.
\end{equation}
Since $\bar\bu_\rmp$ is a solenoidal two-dimensional velocity field we
can write it in terms of a streamfunction $\psi(y,z,t)$:
\begin{equation}
  \bar\bu_\rmp=\bnabla\times(\psi\,\be_x).
\end{equation}
The associated streamwise vorticity is
\begin{equation}
  \omega=-\Delta\psi.
\end{equation}
We therefore replace equations (\ref{uxbar}) and (\ref{upbar}), taking
the curl of the latter to eliminate $\bar p$, with
\begin{equation}
  \p_t\bar u-\f{\p(\psi,\bar u)}{\p(y,z)}+\bar v=
  \epsilon\Delta\bar u+F_x,
\end{equation}
\begin{equation}
  \p_t\omega-\f{\p(\psi,\omega)}{\p(y,z)}=\epsilon\Delta\omega+G_x.
\end{equation}
The reason that streamwise-independent solutions cannot be
self-sustaining is that the streamwise vorticity $\omega$ satisfies an
advection--diffusion equation with no source term (since $\bG$
originates from the streamwise-dependent wave parts of the solution).
For a self-sustaining solution, the mean and wave parts must interact
cooperatively.

\subsubsection{Asymptotic reduction\label{sspnorot}}
Remembering that the streamwise velocity field is strongly amplified by
the lift-up effect and is consequently far larger than the poloidal
velocity field, we seek steady asymptotic solutions in the limit of
very large Reynolds numbers ($\epsilon\ll 1$), of the form
\begin{equation}
\label{asymp1}
  \bar\bu=\bar u_{0}(y,z)\,\be_x+\epsilon\bar\bu_1(y,z)+\cdots,
\end{equation}
\begin{equation}
  \bu'=\epsilon\bu_1'(x,y,z)+\cdots,
\end{equation}
\begin{equation}
  p'=\epsilon p_1'(x,y,z)+\cdots.
\end{equation}
At first order, these are required to satisfy
\begin{equation}
  -\f{\p(\psi_1,\bar u_{0})}{\p(y,z)}+\p_z\psi_1=\Delta\bar u_{0},
\end{equation}
\begin{equation}
  (y+\bar u_{0})\p_x\bu_1'+
  (\bu_1'\cdot\bnabla\bar u_{0}+v_{1}')\,\be_x=-\bnabla p_1'.
\label{u1'}
\end{equation}
At second order, we have
\begin{equation}
  -\f{\p(\psi_1,\omega_1)}{\p(y,z)}=\Delta\omega_1+G_{x2},
\end{equation}
with
\begin{equation}
  \omega_1=-\Delta\psi_1,
\end{equation}
\begin{equation}
  \bnabla\cdot\bu_1'=0,
\end{equation}
\begin{equation}
  \bG_2=-\bnabla\times\overline{\bu_1'\cdot\bnabla\bu_1'}.
\end{equation}
These equations are closed.  Equation~(\ref{u1'}) for the wave is
linear and can be solved by Fourier analysis in $x$.  All the
equations are therefore reduced to a two-dimensional form.
This approach leads to a simplified set of equations which are just a
reduced version of the full problem of determining exact
self-sustaining solutions.  Since the wave involves a single Fourier
harmonic, in some sense we have achieved a radical spectral truncation
of the three-dimensional problem, yet this is fully supported by the
asymptotic analysis.  The two main advantages of this set of equations
are that (i) the equations are two-dimensional; (ii) there are no
small parameters and therefore no small scales to resolve, since the
Reynolds number has been scaled out.  The amplitude of the wave is
arbitrary since it satisfies a homogeneous linear equation.  For a
domain of fixed size, the amplitude of the wave needs to be chosen so
that $\bar u_{0}$ has the right amplitude to give rise to a marginally
stable wave. Alternatively, the amplitude of the wave can be left as a free
parameter and the streamwise wavenumber (or periodicity length)
adjusted to allow the marginally stable wave to exist.

\begin{figure}[ht]
\resizebox{\hsize}{!}{%
\includegraphics[width=10cm]{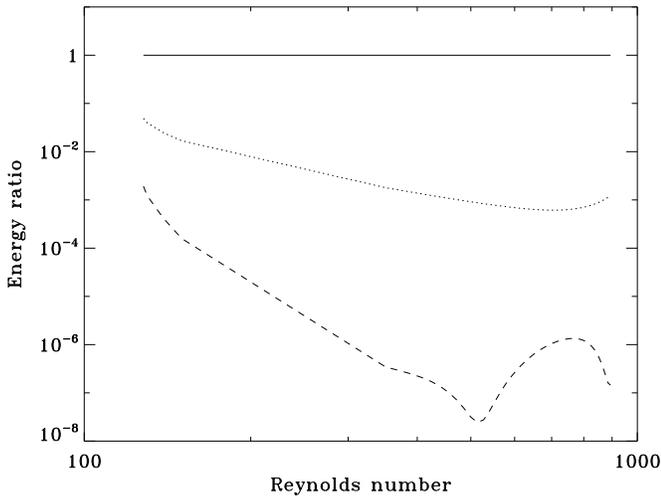}}
   \caption{Vertical ($z$) kinetic energy ratio between either
     $(k_x,k_z)=(\alpha,\beta)$ (full line), $(k_x,k_z)=(2\alpha,\beta)$
     (dotted line) or $(k_x,k_z)=(3\alpha,\beta)$ (dashed line) modes and
     the $(k_x,k_z)=(\alpha,\beta)$ mode ($y=0.707$) for the lower
     branch of non-rotating self-sustaining  solutions, as a function of
     $\rey$. The bump obtained for $\rey>500$ and  $k_x=3\alpha$ can
     be explained by the fact that an error tolerance of $10^{-6}$ for
     the energy was allowed in these computations and that
     the $(N_y,N_z)=(24,32)$ resolution starts to be insufficient for
     such values of $\rey$.}
    \label{fig:evolspec}
 \end{figure}

\begin{figure}[ht]
\resizebox{\hsize}{!}{%
\includegraphics[width=10cm]{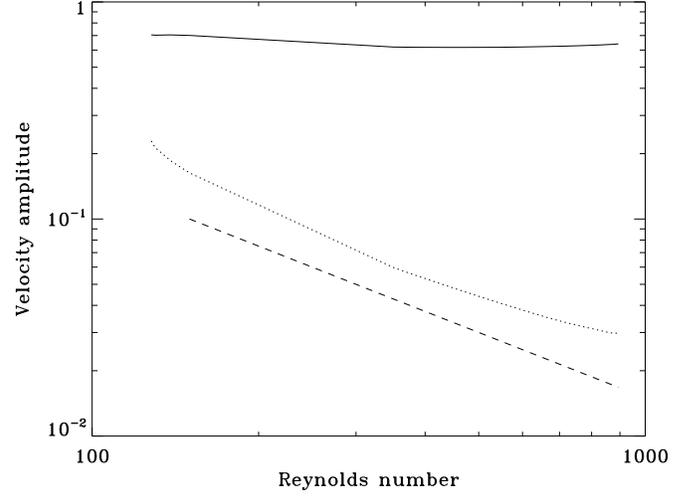}}
   \caption{Maximum of streamwise velocity perturbations $u$ (full line)
     and of poloidal velocity perturbations $\sqrt{v^2+w^2}$ (dotted line)
     for the lower branch of non-rotating self-sustaining solutions, as
     a function of $\rey$. The dashed line corresponds to a function
     proportional to $1/\rey$.}
    \label{fig:maxvel}
 \end{figure}

The agreement between this analysis and the self-sustaining process 
in non-rotating plane Couette flow is demonstrated in
Fig.~\ref{fig:evolspec} and Fig.~\ref{fig:maxvel}.
Figure~\ref{fig:evolspec} shows the evolution with respect to $\rey$
of the ratio between the vertical ($z$) kinetic energy contained in the
$(k_x,k_z)=(\alpha,\beta),(2\alpha,\beta),(3\alpha,\beta)$ Fourier modes
and in the $(k_x,k_z)=(\alpha,\beta)$ mode at $y=0.707$, for the
non-rotating self-sustaining solutions of the lower branch of
Fig.~\ref{fig:bifcouette}. At larger $\rey$, the energy contained in the
$k_x>\alpha$ modes becomes negligible compared to the energy contained
in the fundamental $k_x=\alpha$ mode, which confirms the prediction
made in the previous paragraph that the problem can be solved
using only one Fourier harmonic asymptotically. Figure~\ref{fig:maxvel}
shows that the expansion~(\ref{asymp1}) is very well satisfied by the
non-rotating solutions of the lower branch. The streamwise velocity
amplitude remains approximately constant and $\mathcal{O}(1)$ for
increasing $\rey$, while the poloidal velocity amplitude decreases
like $1/\rey$, as expected for a first order quantity in $\epsilon$. 

One strategy for solving the reduced system would be similar to Waleffe's
approach to the full problem.  One can initially set the wave amplitude to
zero and replace the forcing term $G_{x2}$ by an artificial force of
suitable form, then solve the coupled equations for the ``roll'' $\psi_1$
and ``streak'' $\bar u_{0}$ resulting from this forcing and adjust the
wavenumber $\alpha$ to locate a marginally stable wave.  If the
``feedback'' $G_{x2}$ has the same sign and a roughly similar spatial
form to the artificial forcing, then the solution can be continued by
adding a small amount of wave and reducing the artificial
forcing down to zero. If $\alpha$ is fixed in this process, the
amplitude of the wave will be adjusted in a self-consistent
way. However, we have not carried out such a procedure.

\subsection{Asymptotics on the Rayleigh line}
As demonstrated in Fig.~\ref{fig:evolspec} and Fig.~\ref{fig:maxvel},
the asymptotics for $\rom=0$ are very successful in describing how the
actual the self-sustaining process operates in non-rotating plane
Couette flow. Recalling that the non-rotating solutions are connected to
secondary instabilities of Taylor vortices for small unstable rotation
(Fig.~\ref{fig:nagplot}), it is clear that  the asymptotic
expansion~(\ref{asymp1}) is satisfied for $-\rom\ll 1$ as well with the
Coriolis force playing the role of $\bF$, provided that $-\rom\sim
\epsilon^2$ (or equivalently that the Taylor number defined in
Eq.~(\ref{eq:taylor}) is fixed). This shows that considering the
slightly unstable rotating regime allows to capture the
essential properties of the non-rotating self-sustaining process. 

Here, we similarly analyse the properties of the actual Taylor vortex
flow close to the Rayleigh line and its relations with solutions at
$\rom=-1$, and demonstrate why continuation to $\rom<-1$ regimes through
 three-dimensional solutions is not possible in the same way as in
 the $\rom\sim 0$ region in plane Couette flow.
Introducing a critical amount of rotation $\rom=-1$ to study
self-sustaining behaviour on the Rayleigh line, our equations become
\begin{equation}
  (\p_t+y\p_x+\bu\cdot\bnabla)\,\bu+u\,\be_y=-\bnabla p+\epsilon\Delta\bu,
\end{equation}
\begin{equation}
  \bnabla\cdot\bu=0
\end{equation}
(note the $u\,\be_y$ term here instead of the $v\,\be_x$ term in
Eq.~(\ref{eq:nonrotating})). 
Then
\begin{equation}
  (\p_t+\bar\bu_\rmp\cdot\bnabla)\,\bar u=\epsilon\Delta\bar u+F_x,
\end{equation}
\begin{equation}
  (\p_t+\bar\bu_\rmp\cdot\bnabla)\,\bar\bu_\rmp+\bar u\,\be_y=-\bnabla\bar p+
  \epsilon\Delta\bar\bu_\rmp+\bF_\rmp,
\end{equation}
\begin{equation}
  (\p_t+y\p_x+\bar\bu\cdot\bnabla)\,\bu'+\bu'\cdot\bnabla\bar\bu+
  \bu'\cdot\bnabla\bu'-\overline{\bu'\cdot\bnabla\bu'}+u'\,\be_y=\!
  -\bnabla p'+\epsilon\Delta\bu'
\end{equation}
with
\begin{equation}
  \bnabla\cdot\bar\bu_\rmp=\bnabla\cdot\bu'=0.
\end{equation}
Writing $\bar\bu_\rmp=\bnabla\times(\psi\,\be_x)$ and
$\omega=-\Delta\psi$, this time we have
\begin{equation}
  \p_t\bar u-\f{\p(\psi,\bar u)}{\p(y,z)}=\epsilon\Delta\bar u+F_x,
\end{equation}
\begin{equation}
  \p_t\omega-\f{\p(\psi,\omega)}{\p(y,z)}-\p_z\bar u=\epsilon\Delta\omega+
  G_x.
\end{equation}
The reason that streamwise-independent solutions cannot be
self-sustaining is now that the streamwise velocity satisfies an
advection--diffusion equation with no source term.
Recalling that we are looking for a self-sustaining process in the
$\rom=-1$ regime, this time we would like to take advantage of the
linear anti lift-up effect present in this regime to perform the
asymptotic expansion. The anti lift-up effect predicts generation of
streamwise rolls $\mathcal{O}(1/\epsilon)$ larger than the streamwise
velocity field, which led us to consider two possibilities for
asymptotic expansions.

\subsubsection{$\mathcal{O}(\epsilon)$ streamwise rolls}
A natural asymptotic expansion on the Rayleigh line is obtained
by considering the symmetry of 2D-3C Taylor vortices with
respect to $\rom=-0.5$. If $(u,v,w)$ describes a Taylor vortex solution
at  $\rom$, it can be shown that $(-\rom/(\rom+1)\,u,v,w)$ is a Taylor
vortex solution at $\rom'=-(\rom+1)$. For a given Taylor number,
$-\rom(\rom+1)\sim \epsilon^2$, which
means that if $u$ is the streamwise velocity component of a Taylor
vortex at $-\rom\ll 1$, then  the streamwise velocity component of the
corresponding Taylor vortex at $\rom'=-(\rom+1)$ close to -1 is proportional
to $\epsilon^2 u$. For slightly supercritical Taylor numbers and $-\rom\ll
1$, $v$ and $w$ are $\mathcal{O}(\epsilon)$ and $u$ is $\mathcal{O}(1)$.
In the symmetry transformation, the rolls field remains
$\mathcal{O}(\epsilon)$, while the $u$ field becomes
$\mathcal{O}(\epsilon^2)$ close to $\rom=-1$.  Put differently, for
slightly supercritical Taylor numbers, the streamwise velocity 
for $-\rom\ll 1$ is $\mathcal{O}(1/\epsilon)$ larger than the poloidal
velocity, which is consistent with the lift-up effect present in the
limit $\rom=0$, while for $\rom$ slightly larger than -1, the poloidal
velocity field is $\mathcal{O}(1/\epsilon)$ larger than the streamwise
velocity, which is consistent with the anti lift-up effect present in
the limit $\rom=-1$.  Applying this symmetry transformation to
the $\rom=0$ asymptotic expansion~(\ref{asymp1}), we obtain the
following expansion at $\rom=-1$:
\begin{equation}
\label{asymp3}
  \bar\bu=\epsilon\bar\bu_{\rmp1}(y,z)+\epsilon^2\bar\bu_2(y,z)+\cdots.
\end{equation}
Since the amplitude of this 2D-3C flow is small compared to the
amplitude of the background flow, no secondary instabilities can be
expected. In fact, any marginal three-dimensional secondary solution
$(\bu',p')$ must obey
\begin{equation}
  y\,\partial_x\bu'+u'\vec{e}_y=-\grad{p'}
\end{equation}
at leading order in that case. The rolls field is not involved in this
inviscid linear equation, whose only known solutions are simply 2D-3C
marginal Taylor vortices on the Rayleigh line that are not compatible
with the requirement that $\overline{\bu'}=\vec{0}$.

\begin{figure}[ht]
\resizebox{\hsize}{!}{%
 \includegraphics[width=10cm]{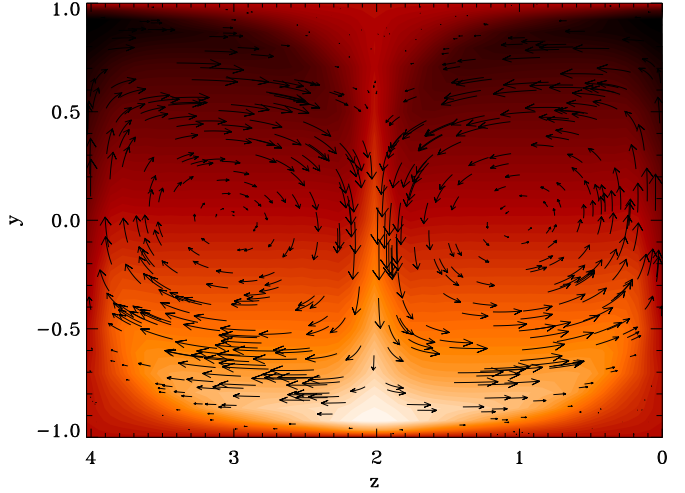}}
   \caption{Streamwise velocity $u$ (color scale) and poloidal $(v,w)$
     velocity perturbations (arrows) of the nonlinear Taylor vortex
     flow for $\beta=1.5585$, $\rom=-0.995$ and $\rey=5000$,
     corresponding to $\tay=1166\,\tay_\rmc$.}
   \label{fig:TV}
\end{figure}

The previous analysis therefore highlights one of the major differences
between small $\rom$ and $\rom\simeq -1$ in rotating plane Couette flow: for
$-\rom\ll 1$, secondary instabilities of the Taylor vortex flow
occur for Taylor numbers close to the critical Taylor number for the
onset of the centrifugal instability because the Taylor vortex flow
is already dominated by $\mathcal{O}(1)$ streaks in the slightly
supercritical regime. For instance, the bifurcation to three-dimensional
solutions in Fig.~\ref{fig:nagplot} occurs for a Taylor number which is
less than two times supercritical.
For $\rom\simeq -1$ and similar weakly supercritical Taylor numbers,
there is no such secondary instability because the Taylor vortex
velocity field is only $\mathcal{O}(\epsilon)$ in that case. The
secondary instabilities computed in Sect.~\ref{numerics_anti}  only appear
at highly supercritical Taylor numbers. For instance, the Taylor number
for which a secondary instability occurs for $\rey=500$ in
Fig.~\ref{fig:bif_WTV} is $\sim 200$ times supercritical. The nature of
these instabilities  remains unclear. They may be related to the presence
of strong velocity gradients in thin shear layers and boundary layers
that form at large Taylor numbers, when nonlinearities (notably advection
of the streamwise velocity by the rolls) become important (see
Fig.~\ref{fig:TV}). Similar conclusions can be drawn when
using the forcing term defined in Eq.~(\ref{eq:forcing}) instead of the
centrifugal approach, because the structure of the corresponding 2D-3C
forced solutions looks very much like the  structure of the nonlinear
Taylor vortices for $\rom$ close to -1.

\subsubsection{$\mathcal{O}(1)$ streamwise rolls\label{ssprot} ?}
As shown previously, an essential requirement for non-rotating
self-sustaining process and three-dimensional dynamics for
$-\rom\ll 1$ is that secondary instabilities of $\mathcal{O}(1)$
structures are present. In the light of these remarks, we finally
attempt to derive an asymptotic expansion for $\rom=-1$ taking advantage
of the anti lift-up effect, that would allow for bifurcations of an
$\mathcal{O}(1)$ $x$-independent velocity field. We seek steady solutions of
the form (see \cite{ogilvie06} for a similar expansion in a different
context):
\begin{equation}
\label{asymp2}
  \bar\bu=\bar\bu_{\rmp0}(y,z)+\epsilon\bar\bu_1(y,z)+
  \epsilon^2\bar\bu_2(y,z)+\cdots,
\end{equation}
\begin{equation}
  \bu'=\epsilon\bu_1'(x,y,z)+\cdots,
\end{equation}
\begin{equation}
  p'=\epsilon p_1'(x,y,z)+\cdots.
\end{equation}
From zeroth to second order, these are required to satisfy
\begin{equation}
\label{eq:zerothorder}
  -\f{\p(\psi_0,\omega_0)}{\p(y,z)}=0,
\end{equation}
\begin{equation}
\label{eq:firstorder}
  -\f{\p(\psi_0,\bar u_{1})}{\p(y,z)}=0,
\end{equation}
\begin{equation}
\label{eq:firstorder2}
  -\f{\p(\psi_0,\omega_1)}{\p(y,z)}-\f{\p(\psi_1,\omega_0)}{\p(y,z)}-
  \p_z\bar u_{1}=\Delta\omega_0,
\end{equation}
\begin{equation}
  (y\p_x+\bar\bu_{\rmp0}\cdot\bnabla)\,\bu_1'+\bu_1'\cdot\bnabla\bar\bu_{\rmp0}+
  u_{1}'\,\be_y=-\bnabla p_1',
\end{equation}
\begin{equation}
  -\f{\p(\psi_0,\bar u_{2})}{\p(y,z)}-\f{\p(\psi_1,\bar u_{1})}{\p(y,z)}=
  \Delta\bar u_{1}+F_{x2},
\end{equation}
with
\begin{equation}
  \omega_0=-\Delta\psi_0,
\end{equation}
\begin{equation}
  \omega_1=-\Delta\psi_1,
\end{equation}
\begin{equation}
  \bnabla\cdot\bu_1'=0,
\end{equation}
\begin{equation}
  \bF_2=-\overline{\bu_1'\cdot\bnabla\bu_1'}.
\end{equation}
Note that a zeroth-order equation must be satisfied here because of the
presence of nonlinear self-interaction terms in the streamwise
vorticity equation, contrary to the $\rom=0$ case where the
zeroth-order terms were strictly vanishing thanks to the
streamwise-indepence of the zeroth-order streaks. 
There is also no equivalent to Eq.~(\ref{eq:firstorder}) in the non-rotating problem.
The meaning of these two zeroth-order and first-order equations is that
advection by the rolls, although it does not affect the overall energy
budget, dominates over viscous and Coriolis terms.

We now drop the subscript on $\psi_0$ and introduce a 
right-handed non-orthogonal coordinate system $(x,\tau,\psi)$ based on the
leading-order streamlines, where $\tau$ is the travel time measured
along a streamline from its intersection with some reference curve.
The Jacobian determinant for this transformation is
$\partial(y,z)/\partial(\tau,\psi)=1$. The preceding equations then
become

\begin{equation}
  \p_\tau\omega_0=0,
\end{equation}
\begin{equation}
  \p_\tau\bar u_{1}=0,
\end{equation}
\begin{equation}
  \p_\tau\omega_1-(\p_\tau\psi_1)\p_\psi\omega_0-
  \p_z\bar
  u_{1}=\p_\tau(g^{\tau\psi}\p_\psi\omega_0)+\p_\psi(g^{\psi\psi}\p_\psi\omega_0),
\end{equation}
\begin{equation}
  \p_\tau\bar u_{2}-(\p_\tau\psi_1)\p_\psi\bar u_{1}=
\p_\tau(g^{\tau\psi}\p_\psi\bar u_{1})+\p_\psi(g^{\psi\psi}\p_\psi\bar u_{1})+F_{x2},
\end{equation}
where $g^{ij}$ are the inverse metric coefficients. The unknown quantities
$\omega_1$ and $\bar u_{2}$ are eliminated by integrating with respect to $\tau$.
Defining the circulation on the streamline labelled by $\psi$ to be
\begin{equation}
  \Gamma(\psi)=\int g^{\psi\psi}\,\rmd\tau,
\end{equation}
 where the integral is along the full length of the streamline,
 the following solvability conditions now need to be satisfied:
\begin{equation}
\label{solv1}
  \f{\rmd}{\rmd\psi}\left(\Gamma\f{\rmd\omega_0}{\rmd\psi}\right)
  +\int(\p_z\bar u_{1})\,\rmd\tau=0,
\end{equation}
\begin{equation}
\label{solv2}
  \f{\rmd}{\rmd\psi}\left(\Gamma\f{\rmd\bar u_{1}}{\rmd\psi}\right)+
  \int F_{x2}\,\rmd\tau=0.
\end{equation}
They state that  the diffusion of the $x$-independent streamwise
vorticity must be balanced by the Coriolis force acting on the
streamwise mean velocity $\bar u_{1}$, while the diffusion of the
$x$-independent streamwise velocity must be balanced by a wave flux (or
by a streamwise force in the absence of three-dimensional
motions). Finally, the first order equations for the ``wave'' read
\begin{equation}
  (y\p_x+\bar\bu_{\rmp0}\cdot\bnabla)\,\bu_1'+\bu_1'\cdot\bnabla\bar\bu_{\rmp0}+
  u_{1}'\,\be_y=-\bnabla p_1',
\end{equation}
\begin{equation}
  \bnabla\cdot\bu_1'=0.
\end{equation}
These equations are  linear and can in principle be solved by Fourier
analysis in $x$, as in the $\rom=0$ case.

To summarize, the asymptotic scenario at $\rom=-1$ can be realized
if i) a particular distribution of streamwise vorticity satisfying
Eq.~(\ref{eq:zerothorder}) 
can be generated
thanks to an artificial forcing term satisfying the solvability
conditions~(\ref{solv1})-(\ref{solv2}) ii) the three-dimensional
instabilities of this forced flow have the ability to replace
the force thanks to their nonlinear feedback on the
streamwise-independent part of the flow. In order to satisfy the first
requirement, it may notably prove necessary to break the point
reflection symmetry 
\begin{equation}
  \label{eq:symTV}
  (y,z)\rightarrow(-y,L_z/2-z),\quad (u,v,w,\omega)\rightarrow(-u,-v,-w,\omega),
\end{equation}
of Taylor vortices which causes the streamwise ``forcing'' term
averaged over closed streamlines in Eq.~(\ref{solv2}) to  vanish.
In the light of the developments presented in this section,
relaxing the constraint of the presence of rigid boundaries that are
responsible for the formation of boundary layers may also help to make
progress. The fact that transition for $\rom\leq 1$ has only
been observed in shearing boxes so far seems to support this view. 
However, we have been unable to isolate such a flow in the course of our
investigations. As mentioned in Sect.~\ref{equations}, standard continuation
methods are not appropriate to compute solutions in shearing boxes
and only theory or direct time-stepping calculations such as those performed by
\cite{hawley99}, \cite{lesur05} or \cite{shen06} may be able to provide
some further insight in that case. An alternative possibility would be
to consider a remapping of the shearwise coordinate to obtain an
unbounded shear flow between $-\infty<y<+\infty$ and to use nonlinear 
continuation techniques for such a flow.

\section{Discussion and conclusions\label{conclusions}}
In this paper, we have tried to build on the current understanding of
transition in non-rotating shear flows in order to isolate exact nonlinear
solutions of the Navier-Stokes equations in Rayleigh-stable
rotating shear flows and to identify a possible route to
turbulence in such flows, including Keplerian flows.
On the one hand, we have shown that nonlinear mechanisms act differently 
on the Rayleigh line, even though analogous transient linear growth
processes exist in both types of flows. We have notably computed exact 
nonlinear solutions for no rotation and for cyclonic rotation but
have not succeeded in isolating solutions for $\rom=-1$ and smaller using
similar techniques. Using the asymptotic approach presented in
Sect.~\ref{theory},  we have shed some light on the differences
between the two regimes and  have demonstrated
quantitatively that one cannot transpose straightforwardly the nonlinear
stability properties of non-rotating shear flows and the phenomenology of
non-rotating self-sustaining processes into the $\rom=-1$ regime. In
particular, our results show that transition in the narrow gap Taylor-Couette
experiment beyond the Rayleigh line, if any, cannot occur through a
self-sustaining mechanism analogous to the one pertaining to
non-rotating plane Couette flow. On the other hand, our findings do not
rule out subcritical transition in Rayleigh-stable shear flows. Finding
a way to satisfy the requirements of
Eqs.~(\ref{eq:zerothorder})-(\ref{solv1})-(\ref{solv2})
may help to discover self-sustaining solutions on the Rayleigh line,
which would then offer an interesting starting point for a possible
continuation towards $\rom=-4/3$. We have argued in Sect.~\ref{theory} that
the presence  of walls may prevent such solutions in plane Couette flow with
$\rom=-1$ and that solutions with the right self-sustaining asymptotic
properties may only be found in unbounded or periodic domains. 

An alternative to a streamwise rolls-based scenario is that the
       transition relies on the existence of shearwise rolls, as suggested by
        \cite{lesur05}. However, such structures are likely to be transient
        due to the ambient shear, and it seems difficult to identify a
        steady self-sustaining process involving them using continuation.
This possibility leads us to finally come back to the Keplerian problem
and to discuss the recent results of \cite{balbus06} and \cite{shen06}
regarding the nonlinear outcome of the transient amplification of
non-axisymmetric shearing waves in this regime.
\cite{balbus06} have notably questioned transition to
turbulence \textit{via} this process by showing that shearing waves
(even three-dimensional ones) do not have any self-interactions and are
therefore exact nonlinear solutions of the equations. We would like to
emphasize that nonlinear interactions of transiently amplified
structures are not a necessary condition for subcritical
transition. Bypass transition in shear flows was highlighted
by \cite{trefethen93}, who argued that this
mechanism could trigger turbulence if initial perturbations could reach
finite amplitude and interact nonlinearly. But a very simple  model for the nonlinear
interactions was used at that time, and \cite{waleffe95}
showed (an argument similar to the one by \cite{balbus06}), using the full
Navier-Stokes equations, that the finite-amplitude streaks generated by
the lift-up effect did not feedback on the streamwise rolls directly,
leaving the outcome of transient growth and the existence of self-sustaining
solutions very uncertain. The real breakthrough of \cite{waleffe95},
however, was to notice that these spanwise modulated structures became
\textit{unstable to three-dimensional 
disturbances} when they reached finite amplitudes, which allowed him
to discover the full self-sustaining process of which this instability
is a crucial piece. In fact, both \cite{balbus06} and \cite{shen06} have
found that the transient amplification of shearing waves leads to fairly
complex dynamical behaviour as well, even though the nonlinear self-interactions
of these structures are vanishing, because such waves become unstable to
Kelvin-Helmholtz instabilities when they reach finite amplitude. These
authors did not observe subsequent self-sustaining behaviour for the
range of parameters explored in their simulations, but more
investigations on this problem may be needed before definite
negative conclusions can be made. 
To be complete, it should also be mentioned that the presence of a
nonlinear saturation mechanism for transiently amplified structures is
not a sufficient condition for subcritical transition either: one of the reasons
why a simple self-sustaining process based on the anti lift-up effect
cannot occur for wall-bounded shear flows in the $\rom=-1$ regime is
precisely the nonlinear advection of the streamwise velocity field by
linearly amplified streamwise rolls. Finally, we emphasize that
following the nonlinear evolution of shearing waves does not ensure that 
the whole nonlinear dynamical behaviour in Rayleigh-stable regimes has been
explored. For instance, the nonlinear cyclonic solutions reported here
do not come out of the transient amplification of \textit{non-axisymmetric} shearing
waves. Instead, they are connected to non-rotating solutions which owe
their existence to the transient growth of \textit{axisymmetric}
structures. Even though we have not been able to isolate nonlinear coherent
structures in the anticyclonic Rayleigh-neutral regime in this study,
solutions in the Keplerian regime, if they exist, may well have a connection
with nonlinear solutions at $\rom=-1$, for which linear axisymmetric
transient growth should be present through the anti lift-up effect. The
numerical results of \cite{lesur05} seem to indicate that transition
at $\rom=-1$ is possible (at least in shearing boxes) and that a
connection with Rayleigh-stable regimes exists, at least for
a limited interval of $\rom$.

Another debated point which has connections with our results is that transition in
disks could not be explained like transition in wall-bounded shear flows
because boundary layers play an important role in transition in the latter case,
whereas they do not exist in accretion disks (or in shearing box simulations).
\cite{balbus06} consider the case of subcritical travelling
Tollmien-Schlichting waves in non-rotating plane Poiseuille flow with
no-slip boundaries to illustrate this idea. Such an argument would be
important  if transition in wall-bounded shear flows was 
known to be triggered by instabilities relying on the presence of
walls. However, as noted in Sect.~\ref{review}, transition in many shear
flows (including plane Poiseuille flow with stress-free boundaries)
cannot be explained this way simply because there is no such instability in these flows.
The self-sustaining process is the most natural explanation available so
far for transition in non-rotating shear flows, but boundary layers are
clearly not an essential part of it. In fact, \cite{waleffe03} has found that 
self-sustaining solutions exist for both no-slip and stress-free
boundary conditions and \cite{lesur05} have shown that transition in
non-rotating or cyclonic shearing boxes and plane Couette flow with
walls occur under very similar conditions.
Also, note that the instability of the streaks field in the
non-rotating case does not result from the shearwise inflection of the mean
velocity profile (which is due to the presence of boundaries at $y=\pm
1$), but has its roots in the  \textit{spanwise} (vertical in the disk
terminology) inflections of the streamwise-independent part of the flow instead.
This kind of inflection is not precluded in flows that are not
wall-bounded. Of course, we have demonstrated in this study that what
happens for Rayleigh-stable anticyclonic regimes is a completely
different story, but  the fact that transition for $\rom\leq -1$ has only
been observed for shearing box configurations and the results of
Sect.~\ref{theory} indicate that the presence of walls at least does not
favour subcritical transition in these regimes.


A lot of issues regarding subcritical transition to turbulence in
rotating shear flows remain very unclear. There is currently no
mathematical proof that there is no such route to turbulence for $\rom\leq -1$,
and the results of \cite{hawley99} and \cite{lesur05} indicate that turbulence indeed
exists in this regime in shearing box configurations. If this is true,
some structures must be present in the phase space of the corresponding nonlinear
dynamical system in order to ignite the transition process, and it should be
possible to identify them. We hope that the results reported in this
paper, which have been obtained by combining linear and nonlinear
approaches, will be helpful to achieve this goal.

 \begin{acknowledgements}
 We would like to thank M.~R.~E. Proctor and A. Antkowiak for several
 helpful discussions. F.~R. acknowledges postdoctoral support from the
 Leverhulme Trust and the Isaac Newton Trust.
 \end{acknowledgements}

\bibliographystyle{aa}
\bibliography{6544}

\begin{thebibliography}{75}
\expandafter\ifx\csname natexlab\endcsname\relax\def\natexlab#1{#1}\fi

\bibitem[{{Afshordi} {et~al.}(2005){Afshordi}, {Mukhopadhyay}, \&
  {Narayan}}]{afshordi05}
{Afshordi}, N., {Mukhopadhyay}, B., \& {Narayan}, R. 2005, ApJ, 629, 373

\bibitem[{{Alfredsson} \& {Tillmark}(2005)}]{alfredsson05}
{Alfredsson}, P.~H. \& {Tillmark}, N. 2005, in IUTAM Symposium on
  Laminar-Turbulent Transition and Finite Amplitude Solutions, ed. T.~{Mullin}
  \& R.~R. {Kerswell} (Springer)

\bibitem[{{Andereck} {et~al.}(1983){Andereck}, {Dickman}, \&
  {Swinney}}]{andereck83}
{Andereck}, C.~D., {Dickman}, R., \& {Swinney}, H.~L. 1983, Phys. Fluids, 26,
  1395

\bibitem[{{Andereck} {et~al.}(1986){Andereck}, {Liu}, \&
  {Swinney}}]{andereck86}
{Andereck}, C.~D., {Liu}, S.~S., \& {Swinney}, H.~L. 1986, J. Fluid Mech., 164,
  155

\bibitem[{{Antkowiak} \& {Brancher}(2006)}]{antko06}
{Antkowiak}, A. \& {Brancher}, P. 2006, submitted to J. Fluid Mech.

\bibitem[{{Balbus}(2003)}]{balbus03}
{Balbus}, S.~A. 2003, ARA{\&}A, 41, 555

\bibitem[{{Balbus} \& {Hawley}(1991)}]{balbus91}
{Balbus}, S.~A. \& {Hawley}, J.~F. 1991, ApJ, 376, 214

\bibitem[{{Balbus} \& {Hawley}(1998)}]{balbus98}
{Balbus}, S.~A. \& {Hawley}, J.~F. 1998, Rev. Mod. Phys., 70, 1

\bibitem[{{Balbus} \& {Hawley}(2006)}]{balbus06}
{Balbus}, S.~A. \& {Hawley}, J.~F. 2006, astro-ph/0608429, to appear in ApJ

\bibitem[{{Balbus} {et~al.}(1996){Balbus}, {Hawley}, \& {Stone}}]{balbus96}
{Balbus}, S.~A., {Hawley}, J.~F., \& {Stone}, J.~M. 1996, ApJ, 467, 76

\bibitem[{{Barnes} \& {Kerswell}(2000)}]{barnes00}
{Barnes}, D.~R. \& {Kerswell}, R.~R. 2000, J. Fluid Mech., 417, 103

\bibitem[{{Bech} \& {Andersson}(1996)}]{bech96}
{Bech}, K.~H. \& {Andersson}, H.~I. 1996, J. Fluid Mech., 317, 195

\bibitem[{{Bech} \& {Andersson}(1997)}]{bech97}
{Bech}, K.~H. \& {Andersson}, H.~I. 1997, J. Fluid Mech., 347, 289

\bibitem[{{Bodo} {et~al.}(2005){Bodo}, {Chagelishvili}, {Murante}, {Tevzadze},
  {Rossi}, \& {Ferrari}}]{bodo05}
{Bodo}, G., {Chagelishvili}, G., {Murante}, G., {et~al.} 2005, A{\&}A, 437, 9

\bibitem[{{Brandenburg} \& {Dintrans}(2006)}]{branden06}
{Brandenburg}, A. \& {Dintrans}, B. 2006, A{\&}A, 450, 437

\bibitem[{{Broadbent} \& {Moore}(1979)}]{broadbent79}
{Broadbent}, E.~G. \& {Moore}, D.~W. 1979, Phil. Trans. R. Soc. Lond., 290, 353

\bibitem[{{Butler} \& {Farrell}(1992)}]{butler92}
{Butler}, K.~M. \& {Farrell}, B.~F. 1992, Phys. Fluids, 4, 1637

\bibitem[{{Chagelishvili} {et~al.}(2003){Chagelishvili}, {Zahn}, {Tevzadze}, \&
  {Lominadze}}]{chagelishvili03}
{Chagelishvili}, G.~D., {Zahn}, J.-P., {Tevzadze}, A.~G., \& {Lominadze}, J.~G.
  2003, A{\&}A, 402, 401

\bibitem[{{C}handrasekhar(1960)}]{chandra60}
{C}handrasekhar, S. 1960, {Proc. Natl. Acad. Sci. USA}, 46, 253

\bibitem[{{Clever} \& {Busse}(1992)}]{clever92}
{Clever}, R.~M. \& {Busse}, F.~H. 1992, J. Fluid Mech., 234, 511

\bibitem[{{Craik} \& {Criminale}(1986)}]{craik86}
{Craik}, A.~D.~D. \& {Criminale}, W.~O. 1986, {Proc. R. Soc. Lond. A}, 406, 13

\bibitem[{{Donati} {et~al.}(2005){Donati}, {Paletou}, {Bouvier}, \&
  {Ferreira}}]{donati05}
{Donati}, J.-F., {Paletou}, F., {Bouvier}, J., \& {Ferreira}, J. 2005, Nature,
  438, 466

\bibitem[{{Drissi} {et~al.}(1999){Drissi}, {Net}, \& {Mercader}}]{drissi99}
{Drissi}, A., {Net}, M., \& {Mercader}, I. 1999, Phys. Rev. E, 60, 1781

\bibitem[{{Dubrulle} {et~al.}(2005){Dubrulle}, {Mari{\'e}}, {Normand},
  {Richard}, {Hersant}, \& {Zahn}}]{dubrulle05}
{Dubrulle}, B., {Mari{\'e}}, L., {Normand}, C., {et~al.} 2005, A{\&}A, 429, 1

\bibitem[{{Faisst} \& {Eckhardt}(2003)}]{faisst03}
{Faisst}, H. \& {Eckhardt}, B. 2003, Phys. Rev. Lett., 91, 224502

\bibitem[{{Fromang} {et~al.}(2002){Fromang}, {Terquem}, \&
  {Balbus}}]{fromang02}
{Fromang}, S., {Terquem}, C., \& {Balbus}, S.~A. 2002, MNRAS, 329, 18

\bibitem[{{Gammie} \& {Menou}(1998)}]{gammie98}
{Gammie}, C.~F. \& {Menou}, K. 1998, ApJ, 492, L75

\bibitem[{{Garaud} \& {Ogilvie}(2005)}]{garaud05}
{Garaud}, P. \& {Ogilvie}, G.~I. 2005, J. Fluid Mech., 530, 145

\bibitem[{{Hamilton} {et~al.}(1995){Hamilton}, {Kim}, \&
  {Waleffe}}]{hamilton95}
{Hamilton}, J.~M., {Kim}, J., \& {Waleffe}, F. 1995, J. Fluid Mech., 287, 317

\bibitem[{{Hawley} {et~al.}(1999){Hawley}, {Balbus}, \& {Winters}}]{hawley99}
{Hawley}, J.~F., {Balbus}, S.~A., \& {Winters}, W.~F. 1999, ApJ, 518, 394

\bibitem[{{Herbert}(1976)}]{herbert76}
{Herbert}, T. 1976, in LNP Vol. 59: Some Methods of Resolution of Free Surface
  Problems, ed. A.~I. {van de Vooren} \& P.~J. {Zandbergen}, 235--240

\bibitem[{{Hof} {et~al.}(2004){Hof}, {van Doorne}, {Westerweel}, {Nieuwstadt},
  {Faisst}, {Eckhardt}, {Wedin}, {Kerswell}, \& {Waleffe}}]{Hof04}
{Hof}, B., {van Doorne}, C.~W.~H., {Westerweel}, J., {et~al.} 2004, {Science},
  305, 1594

\bibitem[{{Ji} {et~al.}(2006){Ji}, {Burin}, {Schartman}, \& {Goodman}}]{ji06}
{Ji}, H., {Burin}, M.~J., {Schartman}, E., \& {Goodman}, J. 2006, Nature, 444,
  343

\bibitem[{{Jones}(1981)}]{jones81}
{Jones}, C.~A. 1981, J. Fluid Mech., 102, 249

\bibitem[{{Knobloch}(1984)}]{knobloch84}
{Knobloch}, E. 1984, Geophys. Astrophys. Fluid Dyn., 29, 105

\bibitem[{{Komminaho} {et~al.}(1996){Komminaho}, A., \&
  {Johansson}}]{komminaho96}
{Komminaho}, J., A., L., \& {Johansson}, A.~V. 1996, J. Fluid Mech., 320, 259

\bibitem[{{Korycansky}(1992)}]{korycansky92}
{Korycansky}, D.~G. 1992, ApJ, 399, 176

\bibitem[{{Landahl}(1980)}]{landahl80}
{Landahl}, M.~T. 1980, J. Fluid Mech., 98, 243

\bibitem[{{Lesur} \& {Longaretti}(2005)}]{lesur05}
{Lesur}, G. \& {Longaretti}, P.-Y. 2005, A{\&}A, 444, 25

\bibitem[{{Longaretti}(2002)}]{longaretti02}
{Longaretti}, P.-Y. 2002, ApJ, 576, 587

\bibitem[{{Lord~Kelvin}(1887)}]{kelvin1887}
{Lord~Kelvin}. 1887, {Phil. Mag.}, 24, 188

\bibitem[{{Mukhopadhyay} {et~al.}(2005){Mukhopadhyay}, {Afshordi}, \&
  {Narayan}}]{narayan05}
{Mukhopadhyay}, B., {Afshordi}, N., \& {Narayan}, R. 2005, ApJ, 629, 383

\bibitem[{{Nagata}(1986)}]{nagata86}
{Nagata}, M. 1986, J. Fluid Mech., 169, 229

\bibitem[{{Nagata}(1988)}]{nagata88}
{Nagata}, M. 1988, J. Fluid Mech., 188, 858

\bibitem[{{Nagata}(1990)}]{nagata90}
{Nagata}, M. 1990, J. Fluid Mech., 217, 519

\bibitem[{{Ogilvie}(2003)}]{ogilvie03}
{Ogilvie}, G.~I. 2003, MNRAS, 340, 969

\bibitem[{{Ogilvie} \& {Lubow}(2006)}]{ogilvie06}
{Ogilvie}, G.~I. \& {Lubow}, S.~H. 2006, MNRAS, 370, 784

\bibitem[{{Orr}(1907)}]{orr1907}
{Orr}, W.~M. 1907, {Proc. R. Irish Acad. A.}, 27, 9

\bibitem[{{Orszag}(1971)}]{orszag71}
{Orszag}, S.~A. 1971, J. Fluid Mech., 50, 689

\bibitem[{{Orszag} \& {Patera}(1980)}]{orszag80}
{Orszag}, S.~A. \& {Patera}, A.~T. 1980, Phys. Rev. Lett., 45, 989

\bibitem[{{Reddy} {et~al.}(1998){Reddy}, {Schmid}, {Baggett}, \&
  {Henningson}}]{reddy98}
{Reddy}, S.~C., {Schmid}, P.~J., {Baggett}, J.~S., \& {Henningson}, D.~S. 1998,
  J. Fluid Mech., 365, 269

\bibitem[{{Reynolds}(1883)}]{reynolds83}
{Reynolds}, O. 1883, Phil. Trans. Roy. Soc., 174

\bibitem[{{Richard}(2001)}]{richard01}
{Richard}, D. 2001, Ph.D.~Thesis

\bibitem[{{Richard} \& {Zahn}(1999)}]{richard99}
{Richard}, D. \& {Zahn}, J.~P. 1999, A{\&}A, 347, 734

\bibitem[{{Rogallo}(1981)}]{rogallo81}
{Rogallo}, R.~S. 1981, {Numerical experiments in homogeneous turbulence - NASA
  \textit{Tech. Mem.} TM-81315}

\bibitem[{{Romanov}(1973)}]{romanov73}
{Romanov}, V.~A. 1973, Funct. Anal. Appl., 7, 137

\bibitem[{{Schmid} \& {Henningson}(2000)}]{schmid00}
{Schmid}, P.~J. \& {Henningson}, D.~S. 2000, Stability and transition in shear
  flows (Springer-Verlag, Berlin)

\bibitem[{{Schmiegel}(1999)}]{schmiegel99}
{Schmiegel}, A. 1999, PhD thesis, {Philipps-Universit\"at Marburg}

\bibitem[{{Shen} {et~al.}(2006){Shen}, {Stone}, \& {Gardiner}}]{shen06}
{Shen}, Y., {Stone}, J.~M., \& {Gardiner}, T.~A. 2006, ApJ, 653, 513

\bibitem[{{Taylor}(1923)}]{taylor23}
{Taylor}, G.~I. 1923, Phil. Trans. Roy. Soc. (London) A, 223, 289

\bibitem[{{Taylor}(1936)}]{taylor36}
{Taylor}, G.~I. 1936, Royal Society of London Proceedings Series A, 157, 546

\bibitem[{{Tevzadze} {et~al.}(2003){Tevzadze}, {Chagelishvili}, {Zahn},
  {Chanishvili}, \& {Lominadze}}]{tevzadze03}
{Tevzadze}, A.~G., {Chagelishvili}, G.~D., {Zahn}, J.-P., {Chanishvili}, R.~G.,
  \& {Lominadze}, J.~G. 2003, A{\&}A, 407, 779

\bibitem[{{Tillmark} \& {Alfredsson}(1996)}]{tillmark96}
{Tillmark}, N. \& {Alfredsson}, P.~H. 1996, in Advances in Turbulence VI, ed.
  S.~{Gavrilakis}, L.~{Machiels}, \& P.~A. {Monkewitz} (Kluwer)

\bibitem[{{Trefethen} {et~al.}(1993){Trefethen}, {Trefethen}, {Reddy}, \&
  {Driscoll}}]{trefethen93}
{Trefethen}, L.~N., {Trefethen}, A.~E., {Reddy}, S.~C., \& {Driscoll}, T.~A.
  1993, Science, 261, 578

\bibitem[{{Umurhan}(2006)}]{umurhan06}
{Umurhan}, O.~M. 2006, MNRAS, 365, 85

\bibitem[{{Velikhov}(1959)}]{velikhov59}
{Velikhov}, E.~P. 1959, Sov. Phys. JETP, 36, 1398

\bibitem[{{Waleffe}(1995)}]{waleffe95}
{Waleffe}, F. 1995, Studies in Applied Math., 95, 319

\bibitem[{{Waleffe}(1997)}]{waleffe97}
{Waleffe}, F. 1997, Phys. Fluids, 9, 883

\bibitem[{{Waleffe}(2001)}]{waleffe01}
{Waleffe}, F. 2001, J. Fluid Mech., 435, 93

\bibitem[{{Waleffe}(2003)}]{waleffe03}
{Waleffe}, F. 2003, Phys. Fluids, 15, 1517

\bibitem[{{Wedin} \& {Kerswell}(2004)}]{wedin04}
{Wedin}, H. \& {Kerswell}, R.~R. 2004, J. Fluid Mech., 508, 333

\bibitem[{{Weidemann} \& {Reddy}(2000)}]{weidemann00}
{Weidemann}, J.~A.~C. \& {Reddy}, S.~C. 2000, {ACM Transactions on Mathematical
  Software}, 26, 465

\bibitem[{{Wendt}(1933)}]{wendt33}
{Wendt}, F. 1933, {Ing. Arch.}, 4, 577

\bibitem[{{Yecko}(2004)}]{yecko04}
{Yecko}, P.~A. 2004, A{\&}A, 425, 385

\bibitem[{{Zahn} {et~al.}(1974){Zahn}, {Toomre}, {Spiegel}, \&
  {Gough}}]{zahn74}
{Zahn}, J.~P., {Toomre}, J., {Spiegel}, E.~A., \& {Gough}, D.~O. 1974, J. Fluid
  Mech., 64, 319

\end{thebibliography}

\end{document}